\documentclass[%
aip,pop,numerical,
 amsmath,amssymb,
preprint,%
superscriptaddress,floatfix]{revtex4-1}

\usepackage{graphicx}% Include figure files
\usepackage{dcolumn}% Align table columns on decimal point
\usepackage{bm}% bold math
\usepackage{placeins}
\usepackage{color}

\usepackage{amssymb}
\usepackage{natbib}
\usepackage{amsmath}

\usepackage{subfigure}

\begin{document}

%\title{Collisionless Magnetic Reconnection in Force-Free Current Sheets}
%\title{Two-Dimensional Particle-in-Cell Simulations of Collisionless Magnetic Reconnection for a Force-Free Current Sheet}
\title{Force-free collisionless current sheet models with non-uniform temperature and density profiles}

% \and 
\author{F. Wilson}
\email{fw237@st-andrews.ac.uk}
\author{T. Neukirch}
\affiliation{School of Mathematics and Statistics, University of St Andrews, St Andrews, UK, KY16 9SS.}
\author{O. Allanson}
\affiliation{School of Mathematics and Statistics, University of St Andrews, St Andrews, UK, KY16 9SS.}
\affiliation{Space and Atmospheric Electricity Group, Department of Meteorology, University of Reading, Reading, RG6 6BB, UK.}

\begin{abstract}
We present a class of one-dimensional, strictly neutral, Vlasov-Maxwell equilibrium distribution functions for force-free current sheets, with magnetic fields defined in terms of Jacobian elliptic functions, extending the results of \citet{Abraham-Shrauner-2013} to allow for non-uniform density and temperature profiles. To achieve this, we use an approach previously applied to the force-free Harris sheet by \citet{Kolotkov-2015}. In one limit of the parameters, we recover the model of \citet{Kolotkov-2015}, while another limit gives a linear force-free field. We discuss conditions on the parameters such that the distribution functions are always positive, and give expressions for the pressure, density, temperature and bulk-flow velocities of the equilibrium, discussing differences from previous models. We also present some illustrative plots of the distribution function in velocity space.
\end{abstract}
%\pacs{52.20.-j, 52.25.Xz, 52.55.-s, 52.65.Ff}

\maketitle

\section{Introduction}
\label{sec: intro}
Force-free current sheets, with magnetic fields satisfying
\begin{eqnarray}
\nabla\cdot\textbf{B}&=&0\label{ff1}\\
\nabla\times\textbf{B}&=&\mu_0\textbf{j}\label{ff2}\\
\textbf{j}\times\textbf{B}&=&0\label{ff3},
\end{eqnarray}
are appropriate for plasma modelling in, e.g., the solar atmosphere and planetary magnetospheres (e.g. Refs.~\onlinecite{Bobrova-1979,Kivelson-1995,Marshbook,Tassi-2008,Panov-2011,Wiegelmann-2012,priest_2014,Vasko-2014a,Zelenyi-2016,Akcay-2016, Burgess-2016, Artemyev-2017a,Artemyev-2017b}). Equations (\ref{ff1})-(\ref{ff3}) imply that the current density is parallel to the magnetic field: $\textbf{j}=\alpha(\textbf{r})\textbf{B}$. The case where $\alpha=0$ defines a potential field, and when $\alpha$ is constant we have a linear force-free field. When $\alpha$ varies with the position $\textbf{r}$, the field is referred to as nonlinear force-free.

Such current sheets as described above can play a crucial role in, e.g, magnetic reconnection processes, for which it is often necessary to consider kinetic length scales (e.g. Ref.~\onlinecite{Birn-Priest}), since many astrophysical plasmas are approximately collisionless. To initialise studies of collisionless reconnection, a Vlasov-Maxwell (VM) equilibrium can be used; since current sheets are strongly localised, they are often well described by one-dimensional (1D) VM equilibrium models. The work by \citet{Wilson-2016} was the first example of a study of collisionless reconnection for which an exact nonlinear force-free equilibrium was used in the initial setup, using a distribution function (DF) found by \citet{Harrison-2009a} for the 'force-free Harris' current sheet,
\begin{equation}
\textbf{B}=B_0(\tanh(z/L), \mbox{sech}(z/L), 0).\label{bffhs}
\end{equation} 
Other studies of collisionless reconnection in force-free current sheets have involved the use of approximate force-free equilibria (e.g. Refs.~\onlinecite{Hesse-2005, Liu-2013, Guo-2014, Guo-2015, Zhou-2015, Guo-2016a, Guo-2016b, Fan-2016}) or linear force-free equilibria (e.g. Refs.~\onlinecite{Bobrova-2001,Nishimura-2003,Bowers-2007}).

To find VM equilibrium DFs consistent with force-free current sheets involves solving the VM equations in the opposite order from what is usually done; a magnetic field satisfying Equations (\ref{ff1})-(\ref{ff3}) is specified, and the DFs are then given by the solution of an inverse problem (e.g. Refs.~\onlinecite{Alpers-1969, Channell-1976,Mottez-2003, Allanson-2016}). As such, finding exact force-free VM equilibria is generally a non-trivial task, and this is reflected in the relatively small number of known solutions. Linear force-free VM equilibria have been discussed in, e.g., Refs.~\onlinecite{Moratz-1966,Sestero-1967,Channell-1976,Correa-Restrepo-1993, Attico-1999,Bobrova-2001,Harrison-2009a}. The first solution for a nonlinear force-free field was found by \citet{Harrison-2009b} (see also Ref.~\onlinecite{Neukirch-2009}) for the force-free Harris sheet, and these solutions were later extended by \citet{Kolotkov-2015} to allow for non-uniform density and temperature profiles (with respect to the spatial coordinate). A number of other equilibrium DFs have also been found for this field. \citet{Wilson-2011} found DFs with an arbitrary dependence on the particle energy; \citet{Stark-2012} discussed DFs in the relativistic limit; \citet{Allanson-2015, Allanson-2016} found DFs in terms of infinite sums over Hermite polynomials, with an arbitrarily low plasma beta (in the previous work on the force-free Harris sheet the plasma beta was constrained to be greater than unity); \citet{Dorville-2015} discussed 'semi-analytic' DFs for a magnetic field which includes the force-free Harris sheet as a special case. 

\citet{Abraham-Shrauner-2013} discussed VM equilibria for a nonlinear force-free magnetic field given in terms of Jacobian elliptic functions. This work can be thought of as a generalisation of some of the previous work, to account for both linear and nonlinear force-free equilibria in one model, since, in one limit of the elliptic modulus, the magnetic field becomes the force-free Harris sheet field, and in another limit it becomes a linear force-free field. The DFs discussed give rise to spatially uniform temperature and density profiles, in a similar way to some of the models mentioned above. In this paper, we will extend this class of DFs to include those consistent with non-uniform temperature and density profiles, using a similar approach used by \citet{Kolotkov-2015} for the force-free Harris sheet. As for Abraham-Shrauner's DFs, the new DFs we will discuss include both the linear force-free limit and the force-free Harris sheet limit\cite{Kolotkov-2015}. 

The paper is laid out as follows; in Section \ref{sec:vm_eq}, we outline the background theory of 1D VM equilibria; in Section \ref{sec: as}, we present an overview of the work by \citet{Abraham-Shrauner-2013}; we discuss the extension of this work to include non-uniform temperature and density profiles in Section \ref{sec:modified_df}, and the velocity space structure of the new DFs is discussed in Section \ref{sec:vspace}; we end with a summary in Section \ref{sec:summary}.

\section{1D Vlasov-Maxwell equilibria}
\label{sec:vm_eq}

In line with some of the previous work on 1D VM equilibria (e.g. Refs.~\onlinecite{Harrison-2009a, Harrison-2009b, Neukirch-2009}), we assume that all quantities depend only on the $z$-coordinate, and that the magnetic field, $\textbf{B}=(B_x,B_y,0)$, can be written as the curl of a vector potential, $\textbf{A}=(A_x, A_y, 0)$. We will not repeat all of the details here, but the result of the above assumptions is that the problem reduces to solving Amp\`{e}re's law in the form
\begin{eqnarray}
\frac{\mathrm{d}^2A_x}{\mathrm{d}z^2}&=&-\mu_0\frac{\partial P_{zz}}{\partial A_x}\label{ampx}\\
\frac{\mathrm{d}^2A_y}{\mathrm{d}z^2}&=&-\mu_0\frac{\partial P_{zz}}{\partial A_y}\label{ampy},
\end{eqnarray}
to find $P_{zz}$, which is the $zz$-component of the pressure tensor, defined by
\begin{equation}
P_{zz}(A_x, A_y)=\sum_sm_s\int v_z^2f_s(H_s, p_{xs}, p_{ys})\mathrm{d}^3v,\label{pzz def}
\end{equation}
where we assume that the DFs can be chosen in such a way that they are compatible with strict neutrality (the scalar potential $\phi=0$) \citep{Channell-1976}. Note that we only consider $P_{zz}$ since this is the component of the pressure tensor which is important for the force-balance of the 1D equilibrium. The DFs, denoted by $f_s$, are assumed to be functions of the particle energy, $H_s=m_s(v_x^2+v_y^2+v_z^2)/2$, and the $x$- and $y$-components of the canonical momentum, $\textbf{p}=m_s\textbf{v}+q_s\textbf{A}$, since these are known constants of motion for a time-independent system with spatial invariance in the $x$- and $y$-directions. Once Amp\`{e}re's law has been solved for $P_{zz}$, the DF can be found by solving Eq. (\ref{pzz def}). This is an example of an inverse problem.%, since the unknown $f_s$ is in the integrand in Eq. (\ref{pzz def}). %A possible method for solving the problem involves the use of Fourier/Weierstrass transforms \citep{Channell-1976,Neukirch-2009, Allanson-2015,Allanson-2016}. 

\section{Abraham-Shrauner's model}
\label{sec: as}
In this section we discuss some properties of the the model developed by \citet{Abraham-Shrauner-2013}, in order to give context to the discussion we will present in Section \ref{sec:modified_df}. In Abraham-Shrauner's work, a nonlinear force-free current sheet profile is considered, described by the magnetic field
\begin{equation}
 \textbf{B}=B_0\Big(\mbox{sn}(z/L), \mbox{cn}(z/L),0\Big),\label{AS field}
\end{equation}
where $B_0$ is a constant, $L$ is the current sheet half-thickness, and $\mbox{sn}$ and $\mbox{cn}$ are Jacobian elliptic functions\cite{NIST:DLMF} with the modulus $k$ suppressed (where $0\le k\le 1$). In the limit $k\to0$, $\mbox{sn}(z/L)\to\sin(z/L)$ and $\mbox{cn}(z/L)\to\cos(z/L)$, and so the magnetic field (\ref{AS field}) becomes the linear force-free field $\textbf{B}=B_0\left(\sin(z/L), \cos(z/L),0\right)$. In the limit $k\to1$, $\mbox{sn}(z/L)\to\tanh(z/L)$ and $\mbox{cn}(z/L)\to\mbox{sech}(z/L)$, giving the force-free Harris sheet magnetic field (Eq. (\ref{bffhs})). The vector potential, $\textbf{A}$, used by \citet{Abraham-Shrauner-2013} is given by
\begin{eqnarray}
A_x&=&\frac{B_0L}{k}\left(\mbox{arcsin}\left(k\mbox{sn}(z/L)\right)+\frac{k\pi}{2}\right)\label{ax}\\
A_y&=&\frac{B_0L}{k}\mbox{ln}\left(\frac{k\mbox{cn}(z/L)+\mbox{dn}(z/L)}{1+k}\right)\label{ay},
\end{eqnarray}
where $\mbox{dn}$ is also an elliptic function. This can be seen by using standard integrals \citep{byrd_friedman} 
%\begin{eqnarray}
%\int\mbox{sn}(x)\mathrm{d}x&=&\frac{1}{k}\ln\left(\mbox{dn}(x)-k\mbox{cn}(x)\right)\\
%\int\mbox{cn}(x)\mathrm{d}x&=&\frac{1}{k}\mbox{arcsin}(k\mbox{sn}(x)).
%\end{eqnarray}
and by choosing the integration constants such that, when $k\to1$, ${A_x}\to 2B_0L\mbox{arctan}(e^{z/L})$, $A_y\to-\ln(\cosh(z/L))$ - the vector potential components used in some of the previous work on the force-free Harris sheet (note also that an alternative gauge for $\textbf{A}$ is discussed for the force-free Harris sheet by \citet{Allanson-2016}). %The $k\to1$ limit of $A_x$ above can be shown by using the identity
%\begin{equation}
%\mbox{arcsin}(\tanh{x})+\pi/2=2\mbox{arctan}(e^x).
%\end{equation}

The current density is given by
\begin{equation}
\textbf{j}=\frac{B_0}{\mu_0L}\left(\mbox{sn}(z/L)\mbox{dn}(z/L),\mbox{cn}(z/L)\mbox{dn}(z/L),0\right)=\frac{{\mbox{dn}(z/L)\textbf{B}}}{\mu_0L}\label{asj},
\end{equation}
and so the force-free parameter $\alpha$ is given by
\begin{equation}
\alpha(z)=\frac{\mbox{dn}(z/L)}{\mu_0L}.
\end{equation}
Note that, in the limit $k\to0$, $\mbox{dn}(z/L)\to1$, and so $\alpha$ is constant (the linear force-free case), but is otherwise a function of position (the nonlinear force-free case).

It is assumed that the pressure has the form $P_{zz}(A_x, A_y)=P_{1}(A_x)+P_{2}(A_y)$; Amp\`{e}re's law in the form of equations (\ref{ampx}) and (\ref{ampy}) can then be solved for $P_{zz}$ in terms of the macroscopic parameters, which gives
\begin{eqnarray}
 P_{zz}&=&P_{t1}+P_{t2}\nonumber\\
 &{}&- \frac{B_0^2}{2\mu_0}\Bigg(\frac{3}{2}+\frac{1}{2k^2}\cos\left(\frac{2kA_x}{B_0L}-k\pi\right)-\frac{1}{4}\left(\frac{1}{k}+1\right)^2\exp\left(\frac{2kA_y}{B_0L}\right)\nonumber\\
 &{}&-\frac{1}{4}\left(\frac{1}{k}-1\right)^2\exp\left(-\frac{2kA_y}{B_0L}\right)\Bigg)\label{pzz_macro},
\end{eqnarray}
where $P_{t1}$ and $P_{t2}$ are constants. This expression can then be used in Eq. (\ref{pzz def}) to determine the DF, which can be written in terms of the constants of motion as
\begin{eqnarray}
f_s(H_s, p_{xs}, p_{ys})&=&\frac{n_{0s}e^{-\beta_sH_s}}{\left(\sqrt{2\pi}v_{th,s}\right)^3}\Bigg[a_{0s}-\frac{1}{2k^2}\exp\left(\frac{(1+k^2)u_{ys}^2}{2v_{th,s}^2}\right)\cos\left(k\beta_su_{xs}p_{xs}-k\pi\right)\nonumber\\
&{}&+\frac{1}{4}\left(\frac{1}{k}+1\right)^2\exp\left(\frac{(1-k^2)u_{ys}^2}{2v_{th,s}^2}\right)\exp({k\beta_su_{ys}p_{ys}})\nonumber\\
&{}&+\frac{1}{4}\left(\frac{1}{k}-1\right)^2\exp\left(\frac{(1-k^2)u_{ys}^2}{2v_{th,s}^2}\right)\exp(-{k\beta_su_{ys}p_{ys}})\Bigg],\label{asdf_rewritten}
\end{eqnarray}
where $a_{0s}$ is a dimensionless constant, $u_{xs}$ and $u_{ys}$ are constant parameters with the dimension of velocity, $\beta_s=(k_BT_s)^{-1}$ and $v_{th,s}=(\beta_sm_s)^{-1/2}$. In the limit $k\to1$, this DF takes the form of that discussed in Refs.~\onlinecite{Harrison-2009b, Neukirch-2009} for the force-free Harris sheet. In the opposite limit, i.e. $k\to0$, it takes a general form which is similar to that described in Refs.~\onlinecite{Channell-1976, Attico-1999,Harrison-2009a}, but with a shift in $p_{xs}$ and $p_{ys}$ (this corresponds to a regauging of the vector potential).

Note that a number of relations exist between the parameters of the model, to ensure positivity of the DFs, strict neutrality, and consistency between the microscopic and macroscopic descriptions of the equilibrium (see Ref.~\onlinecite{Abraham-Shrauner-2013} for further details). Using these relations, the equilibrium density, pressure and temperature can be expressed as
\begin{eqnarray}
n&=&n_0\left(a_0+\frac{1}{2}\right)\label{as n}\\
P_{zz}&=&\frac{n_0(\beta_e+\beta_i)}{\beta_e\beta_i}\left(a_0+\frac{1}{2}\right)\\
T&=&\frac{P_{zz}}{n}=\frac{\beta_e+\beta_i}{\beta_e\beta_i}\label{as temp},
\end{eqnarray}
where $a_0$ and $n_0$ are constant parameters that are introduced when the strict neutrality condition ($\phi=0$) is imposed. The expressions (\ref{as n})-(\ref{as temp}) are independent of the elliptic modulus $k$; this can be seen for $P_{zz}$ through the force-balance equation
\begin{equation}
\frac{B^2}{2\mu_0}+P_{zz} = P_T,
\end{equation}
where $P_T$ is the total pressure, since $B^2=\vert\textbf{B}\vert^2=B_0^2$ for the magnetic field (\ref{AS field}), which is independent of $k$. Since, in this case, $P_{zz}=(\beta_e+\beta_i)n/(\beta_e\beta_i)$, it follows that the density and temperature will also be independent of $k$. As can be seen from the expressions (\ref{as n}) and (\ref{as temp}), Abraham-Shrauner's model has density and temperature profiles that are constant across the current sheet, in a similar way to the models discussed in Refs.~\onlinecite{Harrison-2009a, Harrison-2009b, Neukirch-2009, Wilson-2011, Stark-2012, Allanson-2015, Allanson-2016}. In Section \ref{sec:modified_df}, we discuss how the method of \citet{Kolotkov-2015} can be used to extend the model to have spatially non-uniform density and temperature profiles across the current sheet, while still maintaining a constant pressure as is required for a force-free equilibrium (see e.g. Ref.~\onlinecite{Harrison-2009a}). 

\section{Extension to non-uniform temperature/density case}
\label{sec:modified_df}

To extend the model of \citet{Abraham-Shrauner-2013} to have non-uniform temperature and density profiles, we consider a DF of the form
\begin{eqnarray}
  {f}_{s}&=&\frac{n_{0s}\gamma^{3/2}}{\left(\sqrt{2\pi}v_{th,s}\right)^3}\exp({-\gamma\beta_sH_s})\left(a_{0s}+a_{1s}\cos(\gamma k\beta_su_{xs}p_{xs}-k\pi)\right)\nonumber\\
 &{}&+\frac{n_{0s}}{\left(\sqrt{2\pi}v_{th,s}\right)^3}\exp({-\beta_sH_s})\left(b_{0s}+b_{1s}\exp(k\beta_su_{ys}p_{ys})+b_{2s}\exp(-k\beta_su_{ys}p_{ys})\right)\label{modified AS},
\end{eqnarray}
(where $\gamma>0$) i.e. a modification of Abraham-Shrauner's DF. This corresponds to assuming that the $p_{xs}$-dependent population has a different energy dependence than the $p_{ys}$-dependent population, through the factor $\gamma$. We effectively also have two separate constant background populations (through the constants $a_{0s}$ and $b_{0s}$) whose energy dependences differ. These two populations have been included to allow the limit $k\to0$ to exist, and to ensure this we assume that the constants $a_{0s}$ and $b_{0s}$ scale with the elliptic modulus $k$ in the following way;
\begin{eqnarray}
a_{0s}&=&\bar{a}_{0s}+\frac{\gamma}{2k^2}\exp\left(\frac{u_{xs}^2}{2v_{th,s}^2}\right)\label{a0s choice}\\
b_{0s}&=&\bar{b}_{0s}-\frac{1}{2k^2}\exp\left(\frac{u_{xs}^2}{2v_{th,s}^2}\right)\label{b0s choice},
\end{eqnarray}
for constants $\bar{a}_{0s}$ and $\bar{b}_{0s}$. Note that we have defined the constants in this way so that we have a model that works for all $k$ values between 0 and 1, but for finite small $k$ (or large $u_{xs}/v_{th,s}$), the $k$-dependent parts of $a_{0s}$ and $b_{0s}$ can become very large, which may lead to, e.g., a large maximum density, which may not be physically appropriate. If we were only interested in a particular finite small value of $k$, we could redefine the constants to avoid such issues.%as we like for any value of $k$, as long as we ensure that all physical quantities are properly defined.

%\begin{eqnarray}
% n_s&=&n_{0s}\exp\left(\frac{u_{xs}^2}{2v_{th,s}^2}\right)\Bigg((a_{0s}+b_{0s})\exp\left(-\frac{u_{xs}^2}{2v_{th,s}^2}\right)+a_{1s}\exp\left(-\frac{(1+\gamma k^2)u_{xs}^2}{2v_{th,s}^2}\right)\cos\left(\gamma k\beta_s u_{xs}q_sA_x-k\pi\right)\nonumber\\
% &{}&+b_{1s}\exp\left(\frac{k^2u_{ys}^2-u_{xs}^2}{2v_{th,s}^2}\right)\exp\left(k\beta_s u_{ys}q_sA_y\right)+b_{2s}\exp\left(\frac{k^2u_{ys}^2-u_{xs}^2}{2v_{th,s}^2}\right)\exp\left(-k\beta_s u_{ys}q_sA_y\right)\Bigg),
%\end{eqnarray}
By calculating the number density ($n_s=\int f_s\mathrm{d}^3v$) of the modified DF (\ref{modified AS}), and imposing the condition  $\phi=0$ ($n_i(A_x,A_y)=n_e(A_x,A_y))$, we obtain the neutrality relations (\ref{modified_neut1})-(\ref{modified_neut2}) in Appendix \ref{sec:neutrality_app2}. We can then express $n_s=n$ as
\begin{eqnarray}
n(A_x,A_y)&=&n_0[a_0+b_0+a_1\cos\left(\gamma k\beta_su_{xs}q_sA_x-k\pi\right)\nonumber\\
&{}&+b_1\exp\left(k\beta_su_{ys}q_sA_y\right)+b_2\exp\left(-k\beta_su_{ys}q_sA_y\right)]\label{n ax_ay},
\end{eqnarray}
and the pressure can be calculated from the DF through Eq. (\ref{pzz def}) as
\begin{eqnarray}
  P_{zz}&=&n_{0}\frac{\beta_e+\beta_i}{\beta_e\beta_i}\Bigg(\frac{a_{0}}{\gamma}+b_0+\frac{a_{1}}{\gamma}\cos\left(\gamma k\beta_s u_{xs}q_sA_x-k\pi\right)\nonumber\\
 &{}&+b_{1}\exp\left(k\beta_s u_{ys}q_sA_y\right)+b_{2}\exp\left(-k\beta_s u_{ys}q_sA_y\right)\Bigg).\label{new pzz}
\end{eqnarray}
Note the $\gamma^{-1}$ factors appearing in parts of Eq. (\ref{new pzz}), meaning that the pressure is no longer simply a multiple of the density as in Abraham-Shrauner's model. Eq. (\ref{new pzz}) for the pressure can be compared with Eq. (\ref{pzz_macro}) to give the relations (\ref{micromacro1})-(\ref{ux uy gamma}) (see Appendix \ref{sec:neutrality_app2}) between the microscopic and macroscopic parameters. Using these relations, and the neutrality relations in Appendix \ref{sec:neutrality_app2}, the modified DF (\ref{modified AS}) can then be written as
\begin{eqnarray}
f_s&=&\frac{\gamma^{3/2}n_{0s}\exp(-\gamma\beta_sH_s)}{\left(\sqrt{2\pi}v_{th,s}\right)^3}\Bigg(a_{0s}-\frac{\gamma}{2k^2}\exp\left(\frac{(\gamma k^2+1)u_{xs}^2}{2v_{th,s}^2}\right)\cos(\gamma k\beta_su_{xs}p_{xs}-k\pi)\Bigg) \nonumber\\
&{}&+\frac{n_{0s}\exp(-\beta_sH_s)}{\left(\sqrt{2\pi}v_{th,s}\right)^3}\Bigg(\frac{1}{4}\exp\left(\frac{u_{xs}^2-k^2u_{ys}^2}{2v_{th,s}^2}\right)\nonumber\\
&{}&\times\left\{\left(\frac{1}{k}+1\right)^2\exp(k\beta_su_{ys}p_{ys}) + \left(\frac{1}{k}-1\right)^2\exp(-k\beta_su_{ys}p_{ys})\right\} +b_{0s}\Bigg).\label{new df}
\end{eqnarray}
%which can be written in terms of $\textbf{v}$ and $z$ if required, in a similar way to the Abraham-Shrauner DF (see Eq. (\ref{as df z})). 
%Due to the fact that now the first background population and the $p_{xs}$-dependent one have a different temperature from both the second background population and the $p_{ys}$-dependent one, it is not as straightforward to determine conditions on the parameters such that the 
Sufficient conditions for the positivity of the DF (\ref{new df}) across the whole phase space can be derived by assuming that the functions
\begin{eqnarray}
g_{1s}(p_{xs})&=&a_{0s}-\frac{\gamma}{2k^2}\exp\left(\frac{(\gamma k^2+1)u_{xs}^2}{2v_{th,s}^2}\right)\cos(\gamma k\beta_su_{xs}p_{xs}-k\pi)\\
g_{2s}(p_{ys})&=&b_{0s}+\frac{1}{4}\exp\left(\frac{u_{xs}^2-k^2u_{ys}^2}{2v_{th,s}^2}\right)\nonumber\\
&{}&\times\left(\left(\frac{1}{k}+1\right)^2\exp(k\beta_su_{ys}p_{ys}) + \left(\frac{1}{k}-1\right)^2\exp(-k\beta_su_{ys}p_{ys})\right),
\end{eqnarray}
are both positive, and are given by
\begin{eqnarray}
\bar{a}_0&>&\frac{\gamma}{2k^2}\left[\exp\left(\frac{\gamma k^2u_{xs}^2}{2v_{th,s}^2}\right)-1\right]\label{a0 bar cond}\\
\bar{b}_0&>&\frac{1}{2k^2}\left[1 - (1-k^2)\exp\left(-\frac{k^2u_{ys}^2}{2v_{th,s}^2}\right)\right]\label{b0 bar cond},
\end{eqnarray}
where $\bar{a}_0$ and $\bar{b}_0$ are defined in Appendix \ref{sec:neutrality_app2}. Note that these conditions are well defined in the limit $k\to0$. %, giving 
%\begin{eqnarray}
%\bar{a}_0&>&\frac{\gamma^2u_{xs}^2}{4v_{th,s}^2}\\
%\bar{b}_0&>&\frac{1}{2}\left(1+\frac{u_{ys}^2}{2v_{th,s}^2}\right).
%\end{eqnarray}
Since $0\le k\le1$, $\gamma>0$ and the exponential term in Eq. (\ref{a0 bar cond}) has a minimum value of unity, we see that $\bar{a}_0\ge0$. %For $\gamma=1$, we have the condition
%\begin{eqnarray}
%\bar{a}_{0}>-\bar{b}_0+\frac{1}{2k^2}\left(\exp\left(\frac{k^2u_{ys}^2}{2v_{th,s}^2}\right)-(1-k^2)\exp\left(-\frac{k^2u_{ys}^2}{2v_{th,s}^2}\right)\right),\label{fs positive_modified}
%\end{eqnarray}
%which is just a modified version of the condition (\ref{fs positive}).

The new DF (\ref{new df}) describes an equilibrium with non-uniform density and temperature profiles; we can show this by writing them as functions of $z$ using Equations (\ref{ax}), (\ref{ay}), (\ref{a1})-(\ref{a3}) and the definitions of $\bar{a}_0$ and $\bar{b}_0$, which gives
\begin{eqnarray}
n(z)&=&n_0\Bigg[\bar{a}_0+\bar{b}_0+\frac{1}{2}+(\gamma-1)\mbox{sn}^2(z/L)\Bigg]\label{ns}\\
T(z)&=&\frac{P_{zz}}{n}=\frac{\beta_e+\beta_i}{\beta_e\beta_i}\left(\frac{\bar{a}_0}{\gamma}+\bar{b}_0+\frac{1}{2}\right)\left(\bar{a}_0+\bar{b}_0+\frac{1}{2}+\left(\gamma-1\right)\mbox{sn}^2(z/L)\right)^{-1}\label{temp},
\end{eqnarray}
where the uniform value of the pressure is given by
\begin{equation}
P_{zz}=\frac{n_0(\beta_e+\beta_i)}{\beta_e\beta_i}\left(\frac{\bar{a}_0}{\gamma}+\bar{b}_0+\frac{1}{2}\right)\label{pzz},
\end{equation}
which is independent of the modulus $k$ (for the same reasons as discussed in Section \ref{sec: as}), and is similar to the expression found by \citet{Kolotkov-2015} for the force-free Harris sheet. Note, however, that this time the density depends on $k$, due to the introduction of the $\gamma$ factors in the DF (the pressure can no longer be written as $P_{zz}=(\beta_e+\beta_i)n/(\beta_e\beta_i)$ as it can in the uniform temperature model). It can be seen that, for $\gamma=1$, we recover the constant density/temperature case of \citet{Abraham-Shrauner-2013}.

Provided the DF (\ref{new df}) is positive over the whole phase space, then the density, pressure and temperature will also be positive everywhere. Note, however, that the opposite is not true, i.e. a positive density and pressure do not imply a positive DF. We ensure that the DF is positive by choosing parameters in such a way that the conditions (\ref{a0 bar cond}) and (\ref{b0 bar cond}) are satisfied (for both ions and electrons).
\begin{figure}[htp]
%\centering\
\subfigure[]{\scalebox{0.39}{\includegraphics{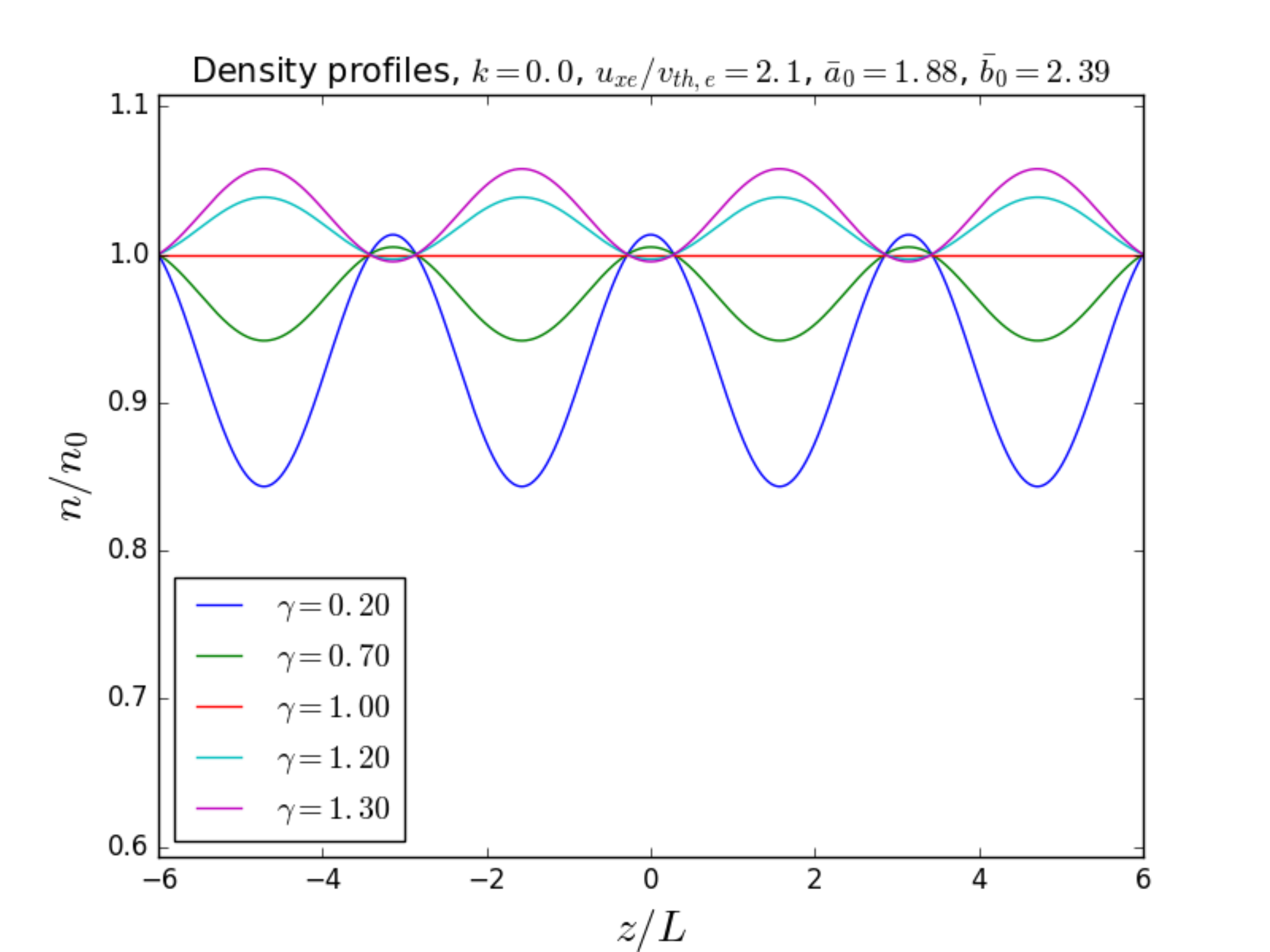}}}
\subfigure[]{\scalebox{0.39}{\includegraphics{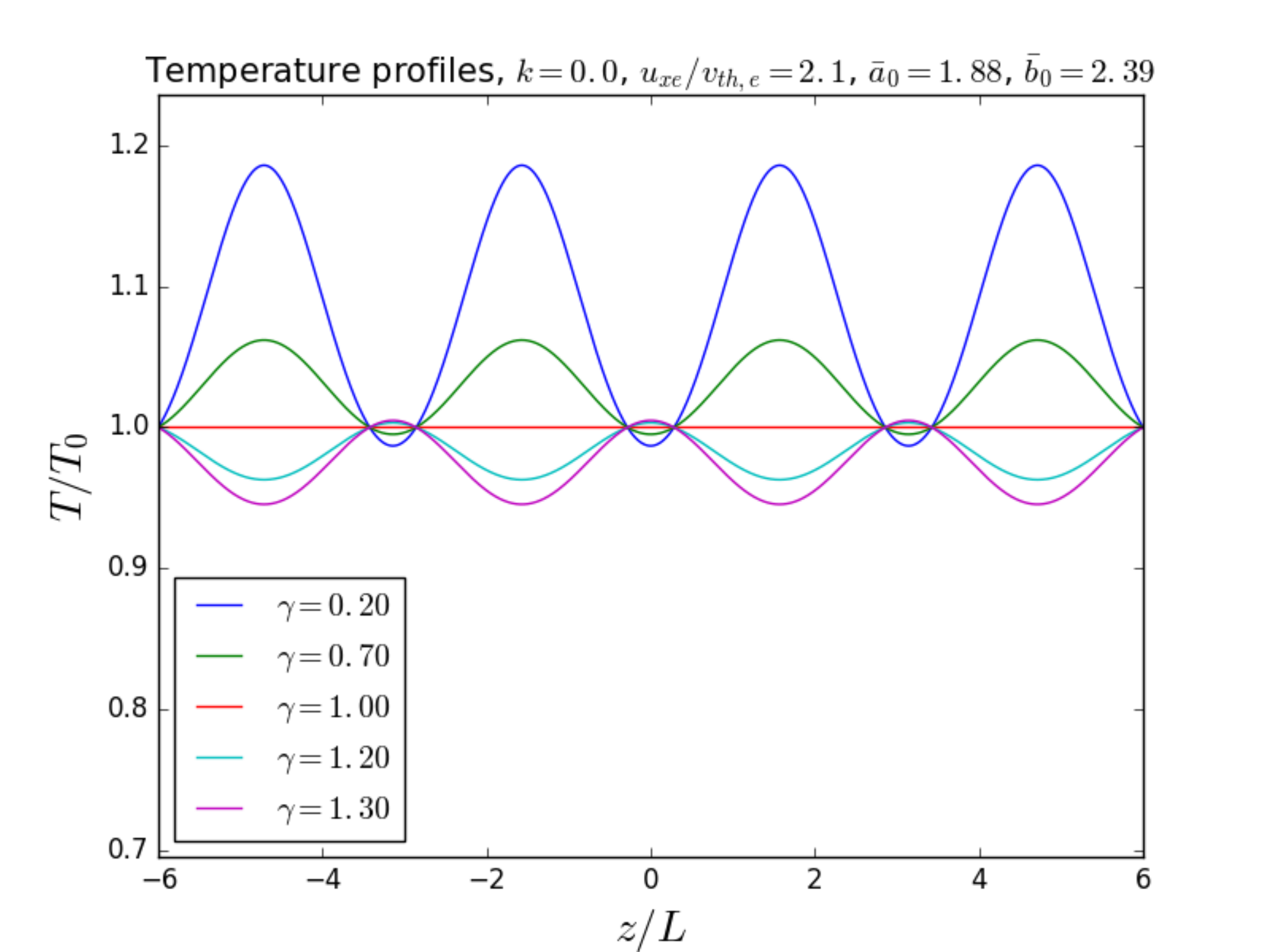}}}
\caption{(a) Density and (b) temperature profiles for various values of $\gamma$, for $k=0$ (the linear force-free case). Both quantities are normalised to have a value of unity at the lower $z$-boundary.}\label{fig:density_temp_ff}
\end{figure}
\begin{figure}[htp]
%\centering\
\subfigure[]{\scalebox{0.39}{\includegraphics{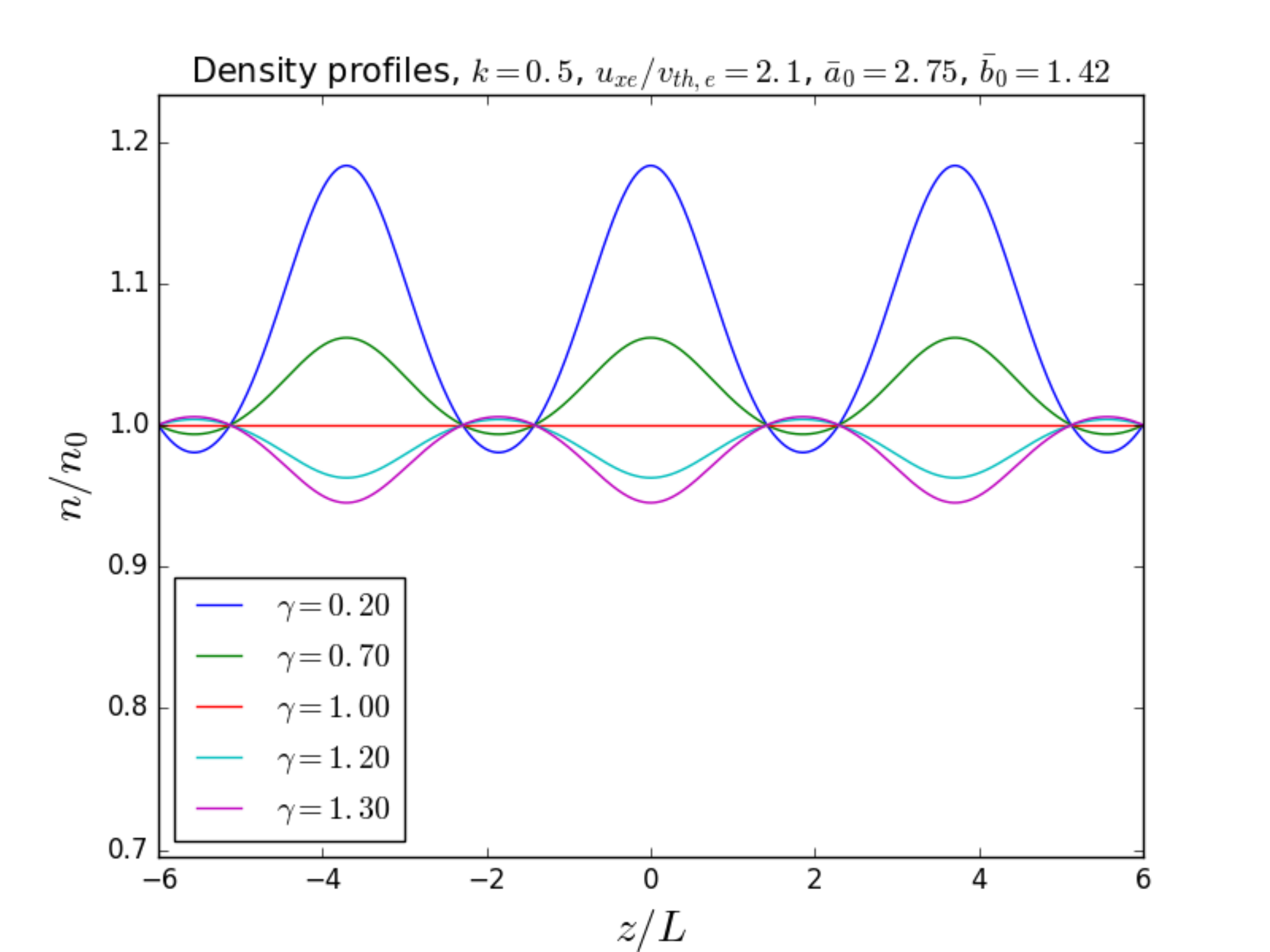}}}
\subfigure[]{\scalebox{0.39}{\includegraphics{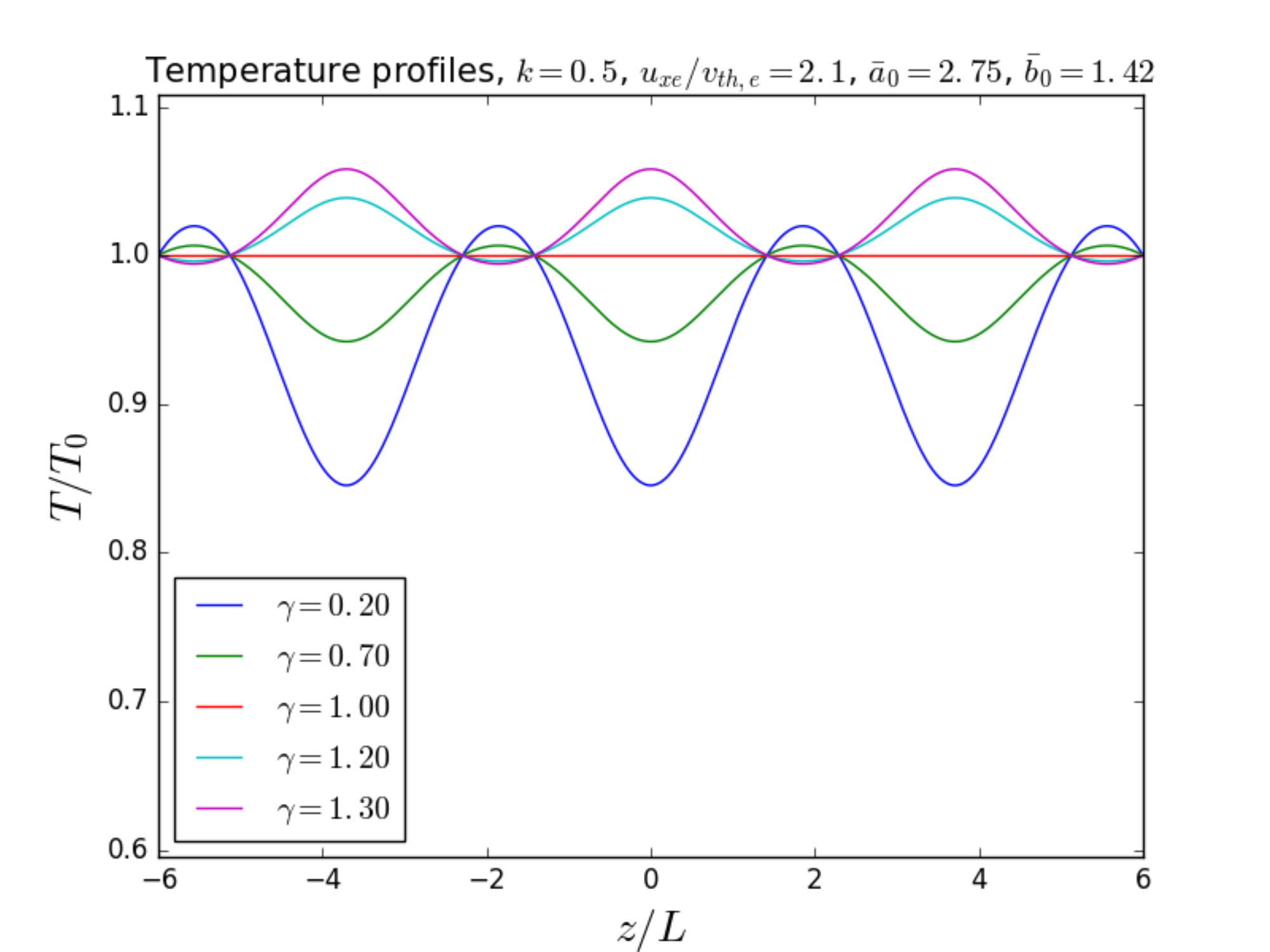}}}
\caption{(a) Density and (b) temperature profiles for various values of $\gamma$, for $k=0.5$. Both quantities are normalised to have a value of unity at the lower $z$-boundary.}\label{fig:density_temp1}
\end{figure}
Figure \ref{fig:density_temp_ff} shows profiles of the density and temperature for different values of $\gamma$, with $k=0$ (the linear force-free case). Figure \ref{fig:density_temp1} shows the same quantities with $k=0.5$. They are normalised to have a value of unity at the lower $z$-boundary of each plot, and we have chosen parameters such that the DFs are positive for ions and electrons (note that if we choose $u_{xe}/v_{th,e}$ then this fixes $u_{xi}/v_{th,i}$ through Eq. (\ref{uxiuxe}), if we specify the mass ratio and the ratio $\beta_e/\beta_i$). For $\gamma=1.0$ in each figure, we see that both the density and temperature are constant, as in Abraham-Shrauner's model. For the other values of $\gamma$ shown, the quantities have a periodic structure. In regions where the density is enhanced/depleted (with respect to the constant value for $\gamma=1$), there is a corresponding depletion/enhancement of the temperature, which ensures that the two quantities multiply together to give a constant pressure, as required for the force-free equilibrium. Additionally, in regions where values of $\gamma>1$ lead to an enhancement/depletion of the quantities, the opposite behaviour is seen when $\gamma<1$, i.e. a depletion/enhancement of the quantities. Similar features are seen by \citet{Kolotkov-2015} (which we obtain in the limit $k\to1$), but note that the density and temperature are not periodic in this case, and so, for a particular $\gamma$ value, there is either an enhancement or depletion of the density/temperature (not both). %We have taken a realistic mass ratio and $\beta_e/\beta_i=1.0$ in both plots - these are needed to check that the ion DFs are positive given the choice of $u_{xe}/v_{th,e}$, which can be expressed as
%\begin{equation}
%\frac{u_{xe}}{v_{th,e}}=\sqrt{\frac{m_i}{m_e}\frac{\beta_i}{\beta_e}}\frac{u_{xi}}{v_{th,i}},
%\end{equation}
%where we have used Eq. (\ref{uxiuxe}).

We will now briefly discuss some other properties of the model. The plasma beta, defined in this case as the ratio of $P_{zz}$ to the magnetic pressure $B_0^2/(2\mu_0)$, is given (using Eq. (\ref{micromacro1})) by%=n_0(\beta_e+\beta_i)/(\beta_e\beta_i)$, is given by
\begin{eqnarray}
\beta_{pl} &=& \frac{\bar{a}_0}{\gamma}+\bar{b}_0+\frac{1}{2}.
\end{eqnarray}
Using the conditions (\ref{a0 bar cond}) and (\ref{b0 bar cond}) for positivity of the DF, we have that
\begin{eqnarray}
\beta_{pl}> \frac{1}{2}+\frac{1}{2k^2}\left[\exp\left(\frac{\gamma k^2u_{xs}^2}{2v_{th,s}^2}\right) - (1-k^2)\exp\left(-\frac{k^2u_{ys}^2}{2v_{th,s}^2}\right)\right].
\end{eqnarray}
For $k=0$ and $k=1$, for example, it is straightforward to show that $\beta_{pl}$ must be greater than unity (as in, e.g., the models in Refs.~\onlinecite{Harrison-2009b, Wilson-2011, Abraham-Shrauner-2013,Kolotkov-2015}), since $u_{xs}^2/v_{th,s}^2\ge0$. 
%\[
 %  \beta_{pl}> 
%\begin{cases}
  %  1/2+(2k^2)^{-1}\left[\exp\left({\gamma k^2u_{xs}^2}/{(2v_{th,s}^2)}\right) - (1-k^2)\exp\left(-{k^2u_{ys}%^2}/{(2v_{th,s}^2)}\right)\right], &  \gamma\ne1\\
%   (2k^2)^{-1}\left[\exp\left({k^2u_{xs}^2}/{(2v_{th,s}^2)}\right) - (1-k^2)\exp\left(-{k^2u_{ys}^2}/{(2v_{th,s}%^2)}\right)\right], & \gamma=1
%\end{cases}\label{n+}
%\]
 %For $P_{zz}>0$, we require
%\begin{equation}
%\bar{a}_0>-\gamma\left(\bar{b}_0+\frac{1}{2}\right).\label{p+}
%\end{equation}
%The positivity of the density $n(z)$ depends on the value of $\gamma$. We have
%\begin{equation}
%\[
 %   \bar{a}_0> 
%\begin{cases}
  %  -(\bar{b}_0+1/2), &  \gamma\ge1\\
    %-(b_0+1/2)+(1-\gamma), & \gamma<1
%\end{cases}\label{n+}
%\]
%\end{equation}
%(using the fact that $0\le\mbox{sn}^2(z/L)\le1$). %The conditions (\ref{p+}) and (\ref{n+}) are always satisfied, using the fact that $\gamma>0$ and that $\bar{a}_0$ and $\bar{b}_0>1/2$ for positivity of the DF.%and so the condition 
%\begin{equation}
%a_0>-\left(b_0+\frac{1}{2}\right)+(1-\gamma)\left(1-\frac{1}{2k^2}\right)
%\end{equation}
%guarantees positivity of the density for all $\gamma$

%Note that, by using a similar argument to that in the previous section, we can see from Eq. (\ref{n ax_ay}) that positivity of the density requires
%\begin{equation}
%a_0>-b_0+a_1-2\sqrt{a_2a_3},
%\end{equation}
%but this does not necessarily imply a positive DF.

The bulk-flow velocity components, defined by
\begin{eqnarray}
\langle{\textbf{V}_s}\rangle&=&\frac{1}{n_s}\int\textbf{v}f_s\mathrm{d}^3v,
\end{eqnarray}
have the form
\begin{eqnarray}
\langle{V_{xs}}\rangle&=&\frac{\gamma u_{xs}\mbox{sn}(z/L)\mbox{dn}(z/L)}{\bar{a}_0+\bar{b}_0+1/2+(\gamma-1)\mbox{sn}^2(z/L)}\\
\langle{V_{ys}}\rangle&=&\frac{u_{ys}\mbox{cn}(z/L)\mbox{dn}(z/L)}{\bar{a}_0+\bar{b}_0+1/2+(\gamma-1)\mbox{sn}^2(z/L)}\\
\langle{V_{zs}}\rangle&=&0.
\end{eqnarray}
Through these expressions, we see the role played by the parameters $u_{xs}$ and $u_{ys}$, which can also be written in terms of the ratio of the species gyroradius, $r_{g, s}$, to the current sheet half-width, $L$, by using Eq. (\ref{ux uy gamma}) (similarly to \citet{Neukirch-2009}) as
\begin{eqnarray}
\frac{u_{ys}^2}{v_{th,s}^2}&=&\frac{\gamma^2u_{xs}^2}{v_{th,s}^2}=4\frac{r_{g,s}^2}{L^2}.
\end{eqnarray}
The current density can be calculated from the bulk flow velocity as
\begin{eqnarray}
\textbf{j}&=&\sum_sq_sn_s\langle\textbf{V}_s\rangle,
\end{eqnarray}
and has components 
\begin{eqnarray}
j_x&=&n_{0}e\gamma(u_{xi}-u_{xe})\mbox{sn}(z/L)\mbox{dn}(z/L)\\
j_y&=&{n_0e}(u_{yi}-u_{ye})\mbox{cn}(z/L)\mbox{dn}(z/L)\\
j_z&=&0.
\end{eqnarray}
Using Equations (\ref{micromacro1}) and (\ref{asL}), we can show that these expressions are equivalent to those obtained macroscopically from Amp\`{e}re's law (Eq. (\ref{asj})). 

In the models in e.g. Refs.~\onlinecite{Harrison-2009b,Neukirch-2009, Abraham-Shrauner-2013}, the spatial structure of the current density is determined solely by the structure of the bulk flow velocity since the density is constant, in contrast to the classic Harris sheet model \citep{Harris-1962}, where the bulk flow velocity is constant, and it is the spatial dependence of the density that determines the structure of the current density. In this extended model (and also that of \citet{Kolotkov-2015}), however, both the bulk-flow velocity and density are spatially dependent, and so the spatial structure of the current density is determined from the product of the two quantities.

\subsection{Limiting values of $k$}
In the limit $k\to1$, the number density, temperature, and pressure (Equations (\ref{ns})-(\ref{pzz})) go to the form discussed by \citet{Kolotkov-2015} for the force-free Harris sheet, and the DF (\ref{new df}) becomes the Kolotkov DF (note that our notation is slightly different).
%For the choice
%\begin{eqnarray}
%\bar{a}_0&=&b-\frac{\gamma}{2}\\
%\bar{b}_0&=&\frac{1}{2},
%\end{eqnarray}
%with $b$ a constant (using Kolotkov's notation), t
%\begin{eqnarray}
%f_s&=&\frac{\gamma^{3/2}n_{0s}\exp(-\gamma\beta_sH_s)}{\left(\sqrt{2\pi}v_{th,s}\right)^3}\Bigg(a_{0s}+\frac{\gamma}{2}\exp\left(\frac{(\gamma+1)u_{xs}^2}{2v_{th,s}^2}\right)\cos(\gamma\beta_su_{xs}p_{xs})\Bigg) \nonumber\\
%&{}&+\frac{n_{0s}\exp(-\beta_sH_s)}{\left(\sqrt{2\pi}v_{th,s}\right)^3}\exp(\beta_su_{ys}p_{ys}),\end{eqnarray}
%Kolotkov uses slightly different notation, but the two cases can be made to match by appropriate choices of $\bar{a}_0$ and $\bar{b}_0$. %, where $a_{0s}=b\exp(u_{xs}^2/(2v_{th,s}^2))$ (Kolotkov's '$b_s$').

In the limit $k\to0$, the field becomes linear force-free, and we get a DF of the form
\begin{eqnarray}
f_s&=&\frac{\gamma^{3/2}n_{0}\exp(-\gamma\beta_sH_s)}{\left(\sqrt{2\pi}v_{th,s}\right)^3}\Bigg(\bar{a}_{0}-\frac{\gamma^2u_{xs}^2}{4v_{th,s}^2}+\frac{\gamma}{4}(\gamma\beta_su_{xs}p_{xs}-\pi)^2\Bigg) \nonumber\\
&{}&+\frac{1}{4}\frac{n_{0}\exp(-\beta_sH_s)}{\left(\sqrt{2\pi}v_{th,s}\right)^3}\left(4\bar{b}_{0}-2-\frac{u_{ys}^2}{v_{th,s}^2}+(\beta_su_{ys}p_{ys}+2)^2\right),\label{k to 0}
\end{eqnarray}
which is a modified form of the DF obtained in the $k\to0$ limit of the DF (\ref{asdf_rewritten}). The density and temperature have the form given by Equations (\ref{ns}) and (\ref{temp}) respectively, where $\mbox{sn}(z/L)=\sin(z/L)$.

%As mentioned in Section \ref{sec: as}, \cite{Channell-1976} (in Example C) discusses a DF of the form
%\begin{equation}
%f_s=e^{-\beta_sH_s}\left(a_{0s}+a_{2s}(p_{xs}^2+p_{ys}^2)\right),
%\end{equation}
%which is also consistent with the linear force-free field discussed throughout this paper. In Appendix \ref{app: channell} we discuss a modification of this DF to allow non-uniform temperature and density profiles.%Note that the DFs (\ref{channell df}) and (\ref{k to 0}) (and Channell's DF in Example C, but he does not mention this) give rise to the restriction $T_e/T_i=m_i/m_e$, so that $n_i=n_e$, which may not be physically appropriate for a realistic value of the mass ratio. \textbf{why don't we have this restriction for the other $k$ cases?}
%\begin{eqnarray}
%n(z)=n_0\Bigg[a_0+b_0+\frac{1}{2}+(\gamma-1)\sin^2(z/L)\Bigg]\label{ns_k0}
%\end{eqnarray}

%\begin{itemize}

%\item The vector potential was gauged in such a way that the FFHS limit was satisfied for $k\to1$. If we used a different gauge we might get the Channell DF? i.e. without a shifted pxs/pys?
%\end{itemize}

\section{Velocity space structure of DF}
\label{sec:vspace}
In this section, we present some illustrative plots of the DF (\ref{new df}) to show the effect of changing $\gamma$, i.e. the effect of changing the energy dependence of the different particle populations. In the $v_x$- and $v_y$- directions, it is possible to choose sets of parameters for which there are multiple peaks in the DF, which may have implications for the stability of the equilibrium. \citet{Neukirch-2009} and \citet{Abraham-Shrauner-2013} derive conditions on the parameters in their models such that their DFs will be single-peaked over the whole phase space. Due to the increased complexity of the DFs in terms of energy dependence, however, we have not yet carried out a full analysis of the velocity space structure - this is left for a future investigation.

In the discussion of the plots below, we will refer to cases where the $p_{xs}$ population is 'hotter'/'colder' than the $p_{ys}$ one. This refers to the $p_{xs}$ population having an energy dependence resulting in a 'narrower'/'wider' Maxwellian factor in the DF than the $p_{ys}$ one. We note, however, that because the DFs are not purely Maxwellian, the temperature cannot be properly defined in terms of the width of the DF, but the widths of the first and second parts of the DF gives us a qualitative measure of the temperature difference between the different populations. This notion of temperature should not be confused with the definition of the temperature given in Eq. (\ref{temp}).

\subsection{$v_x$-direction}
\begin{figure}[htp]
 \centering{\scalebox{0.4}{\includegraphics{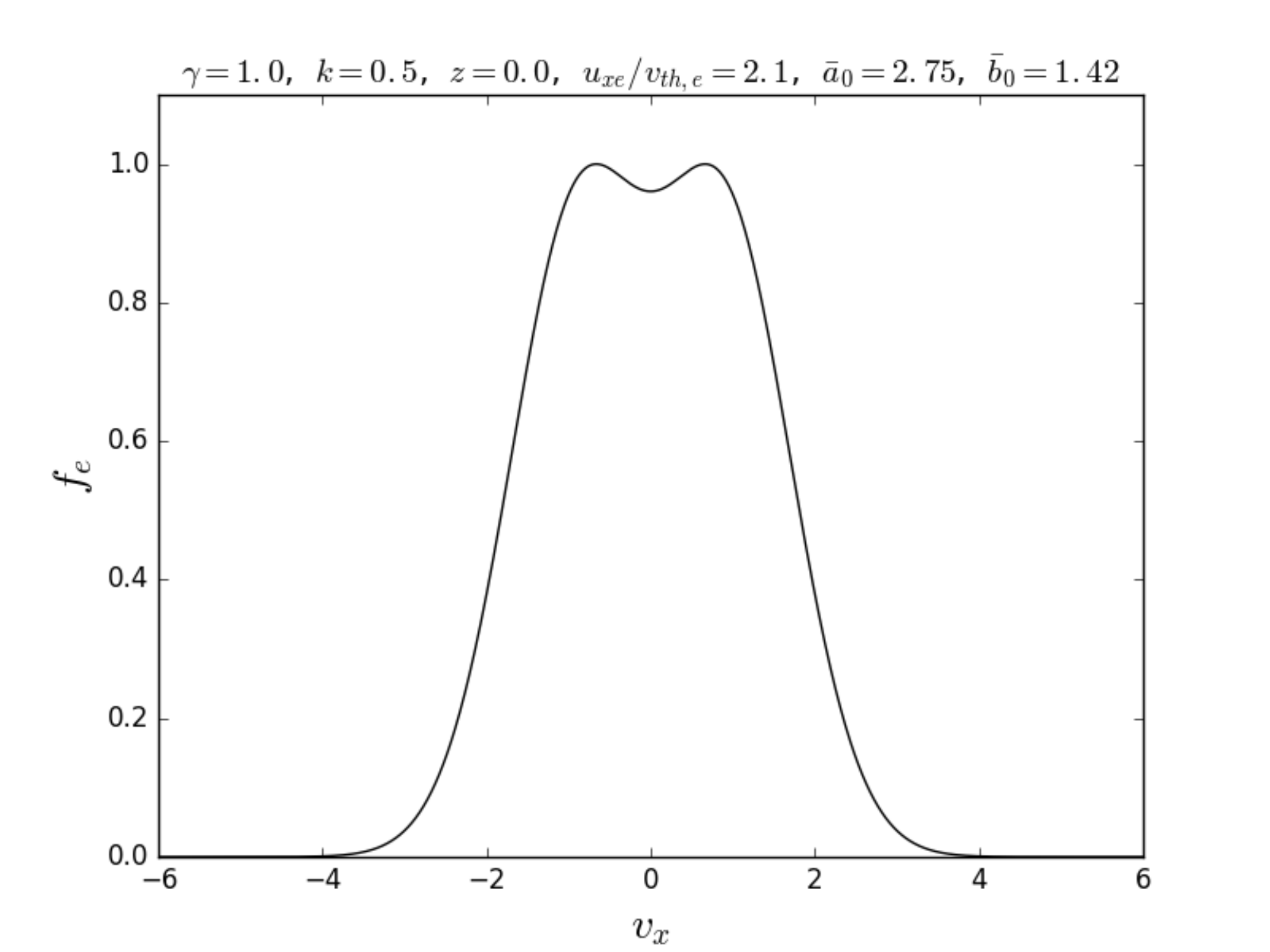}}}
\caption{The electron DF in the $v_x$-direction (with $v_y=v_z=0$) for $\gamma=1$.}\label{fig:gamma1}
\end{figure}

In Figure \ref{fig:gamma1}, we plot the electron DF (\ref{new df}) in the $v_x$-direction (for $v_y=v_z=0$) with $\gamma=1$ (i.e. the Abraham-Shrauner DF). We have chosen a set of parameters for which, at $z=0$, the DF has a double maximum in $v_x$ (these are the same parameters as in Figure \ref{fig:density_temp1}). We note, however, that it is also possible to choose parameters for which the DF has only a single maximum in $v_x$ over the whole phase space, if required (by increasing the density of the background populations appropriately). In Figure \ref{fig:gamma1}, and all subsequent figures in this paper, we normalise the DF to have a maximum value of unity.%, and have taken appropriate values of $\bar{a}_0$ and $\bar{b}_{0}$ such that the DF is positive. 

\begin{figure}[htp]
\centering\
\subfigure[]{\scalebox{0.32}{\includegraphics{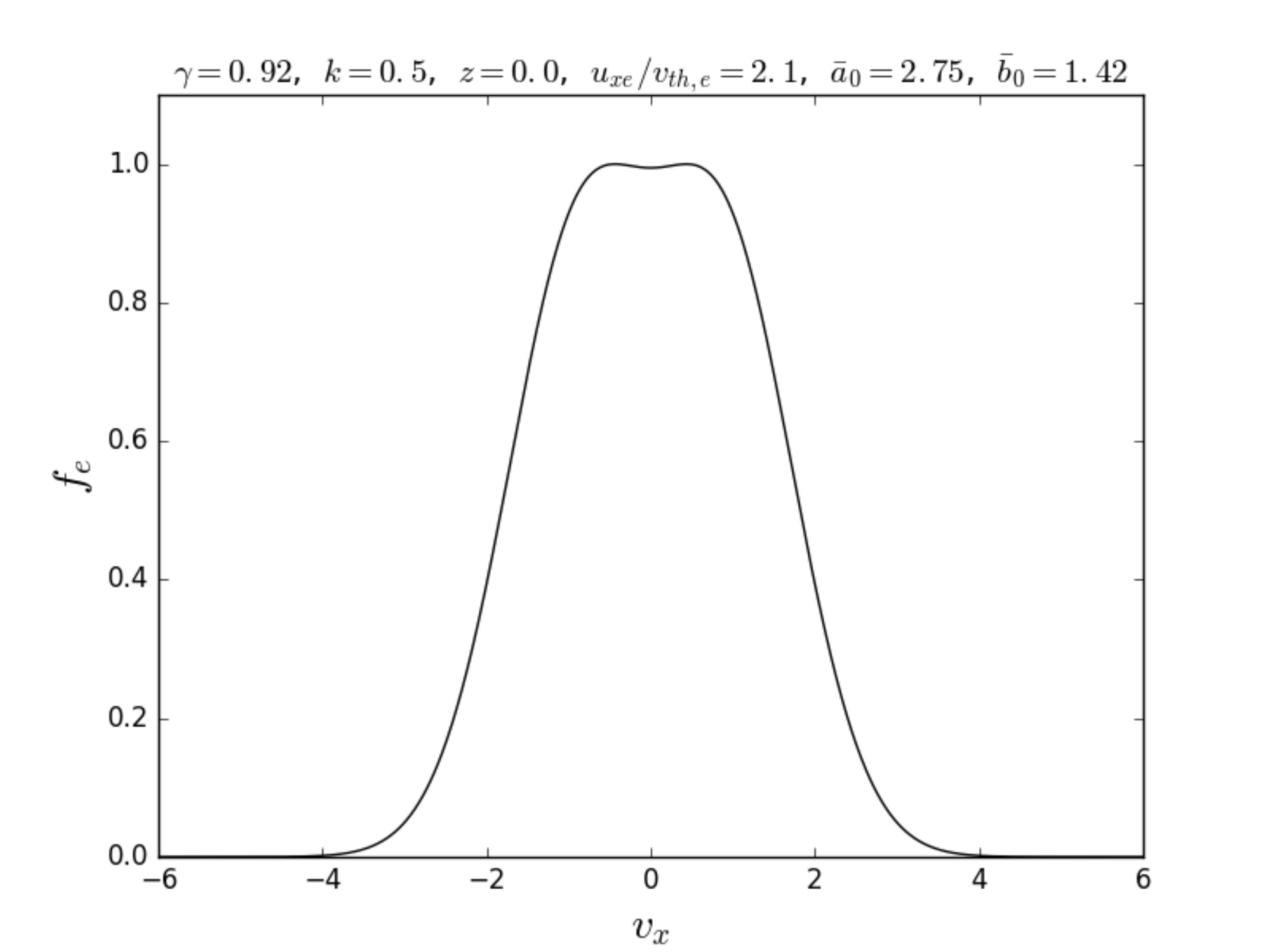}}}
\subfigure[]{\scalebox{0.32}{\includegraphics{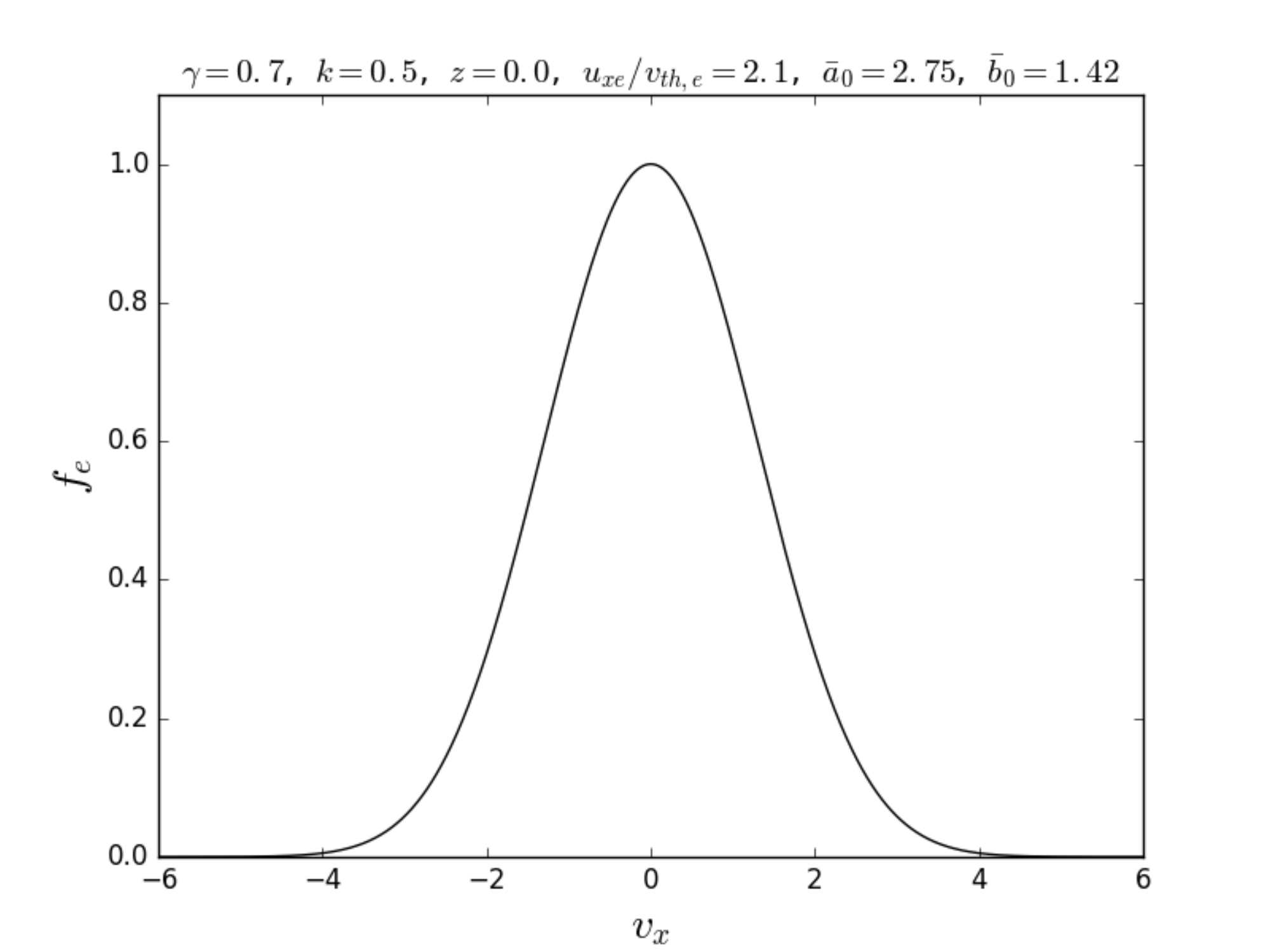}}}
\subfigure[]{\scalebox{0.32}{\includegraphics{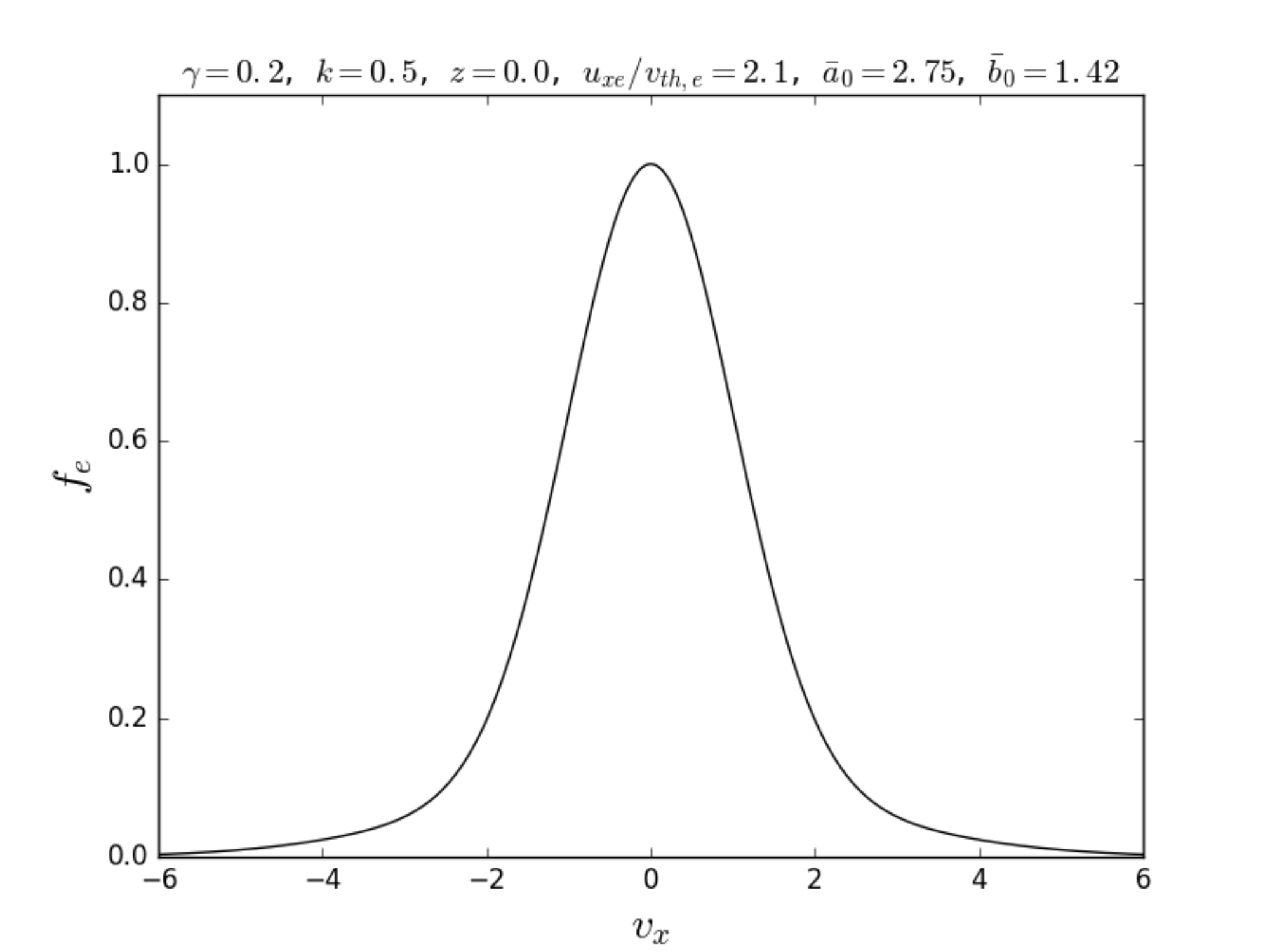}}}
\caption{The electron DF in the $v_x$-direction (with $v_y=v_z=0$) for (a) $\gamma=0.92$, (b) $\gamma=0.7$, (c) $\gamma=0.2$.}\label{fig:gamma_lt1}
\end{figure}

Our main aim in this section is to investigate the effect of changing $\gamma$ on the velocity space structure of the DF. This is why we have chosen parameters that give a double maximum for $\gamma=1$, since the effect of changing $\gamma$ is illustrated more clearly in such cases. Figure \ref{fig:gamma_lt1} shows plots of the electron DF for various values of $\gamma$ which are less than unity.
For $\gamma=0.92$, the double maximum still exists, but has become more slight; for the smaller values of $\gamma$ shown ($0.2$ and $0.7$), the double maximum has disappeared. In the $v_x$-direction, the second part of the DF (which does not depend on $\gamma$) has the Maxwellian form $g(p_{ys})\exp(-\beta_sH_s)$. For $\gamma<1$, the $p_{xs}$-dependent population and the first background one are `hotter` than the $p_{ys}$-dependent and second background populations, and so the Maxwellian factor $\exp(-\gamma\beta_sH_s)$ (in the first part of the DF) has a narrower width than the factor in the second part of the DF. The 'narrow' first part of the DF, including the cosine which can give double maxima in $v_x$, is therefore `swamped` by the wider second part for decreasing $\gamma$, and we see the behaviour in Figure \ref{fig:gamma_lt1}. 

\begin{figure}[htp]
\centering\
\subfigure[]{\scalebox{0.32}{\includegraphics{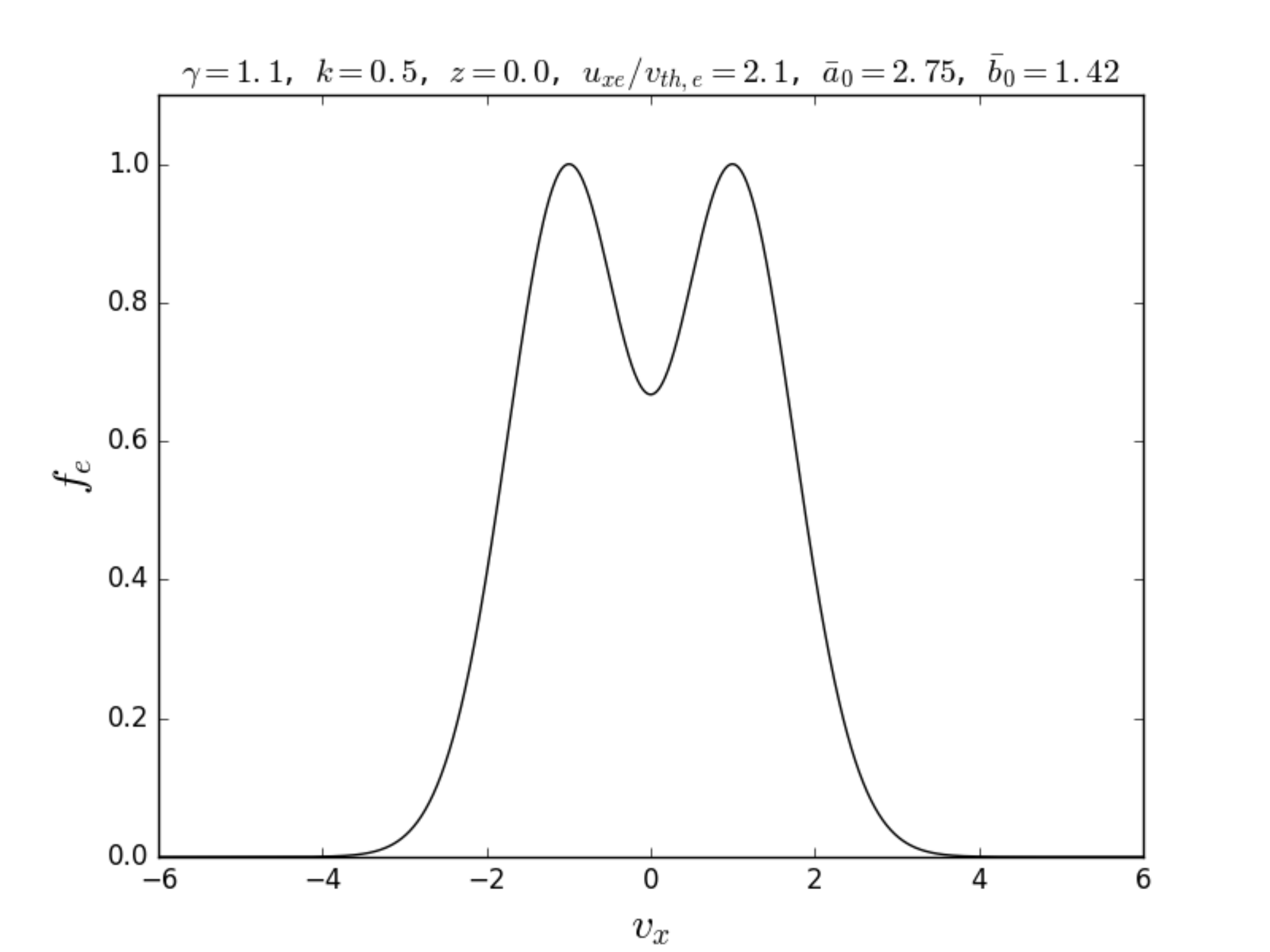}}}
\subfigure[]{\scalebox{0.32}{\includegraphics{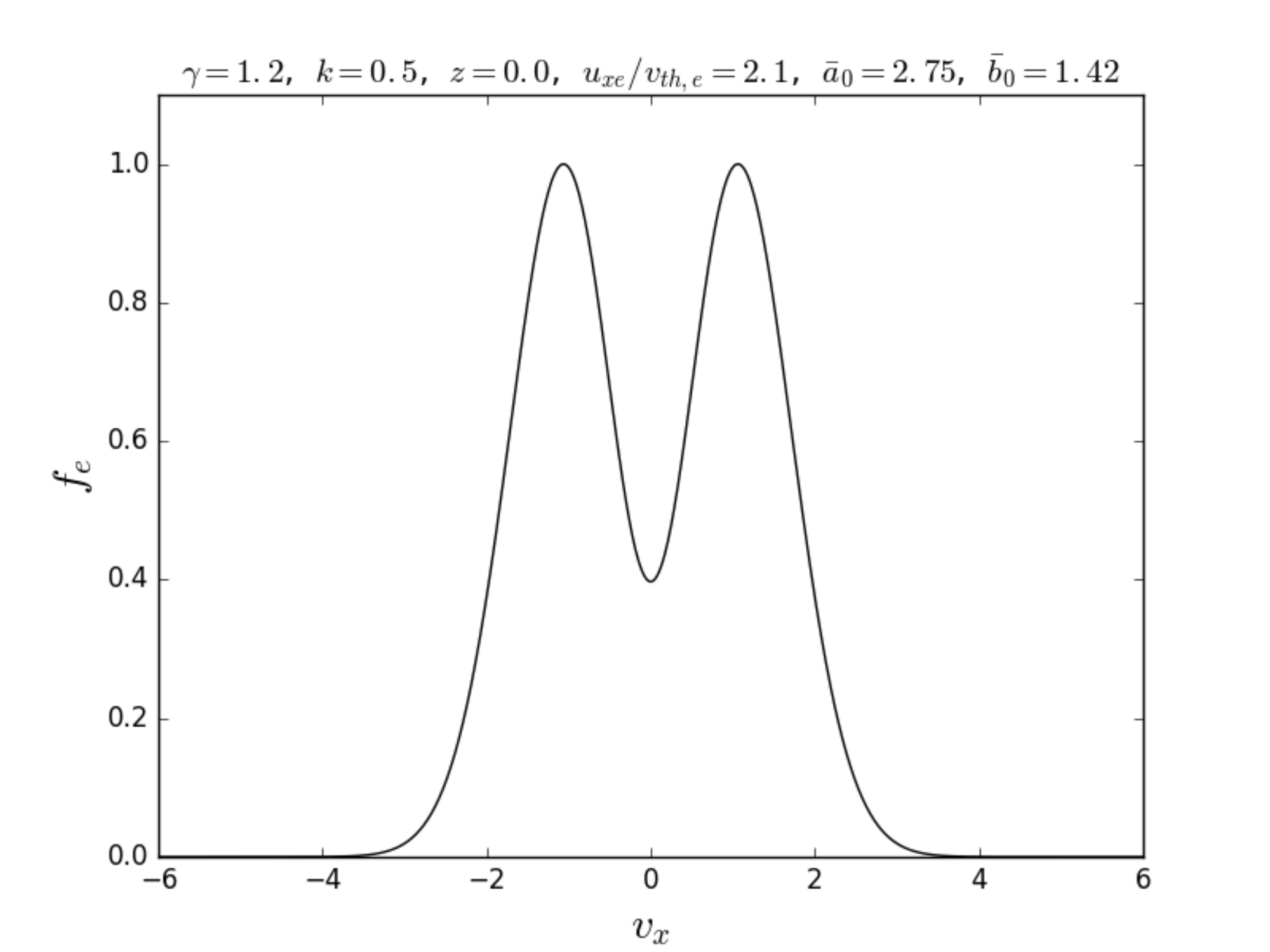}}}
\subfigure[]{\scalebox{0.32}{\includegraphics{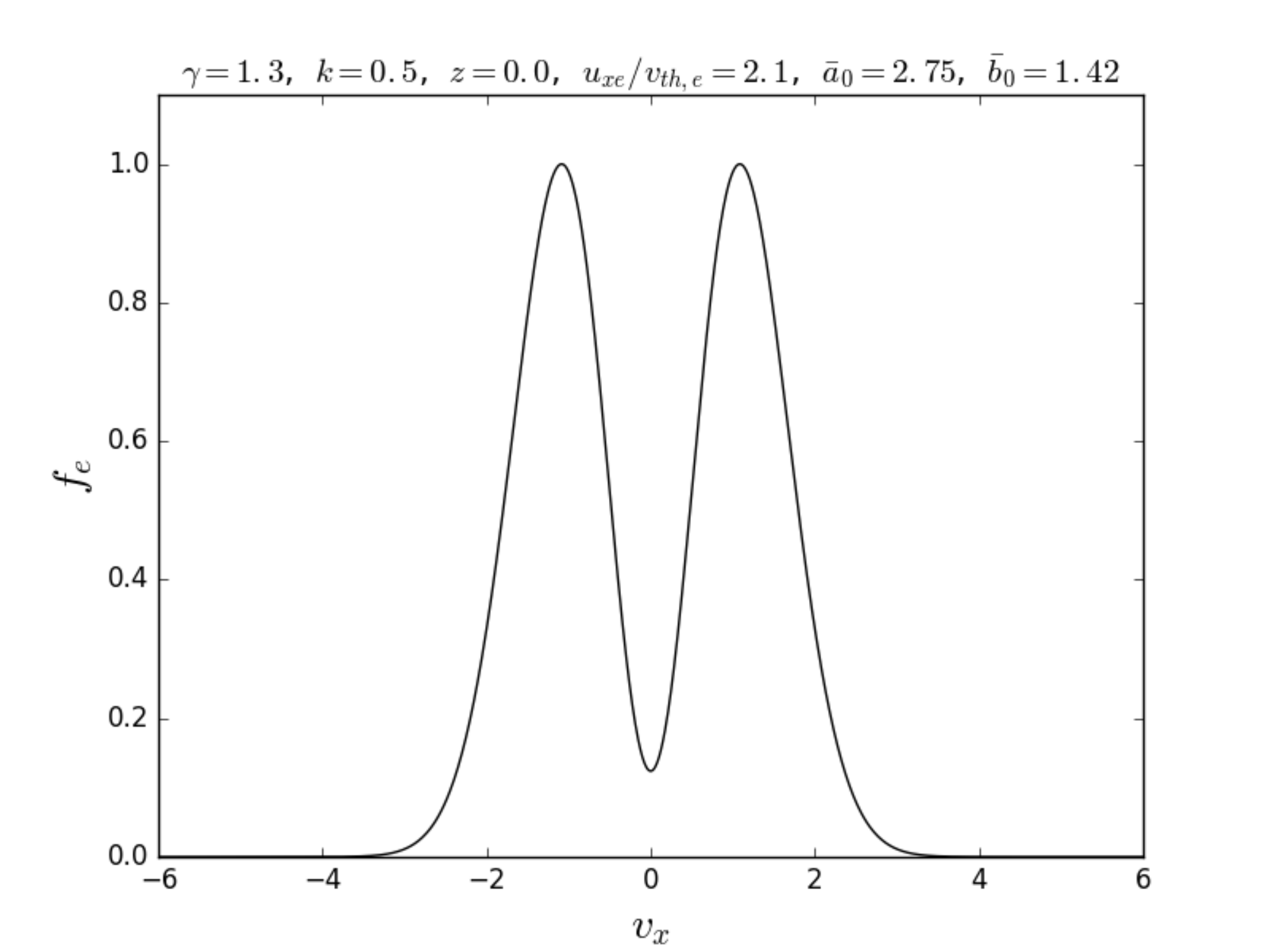}}}
\caption{The electron DF in the $v_x$-direction (with $v_y=v_z=0$) for (a) $\gamma=1.1$, (b) $\gamma=1.2$, (c) $\gamma=1.3$.}\label{fig:gamma_gt1}
\end{figure}

Figure \ref{fig:gamma_gt1} shows plots of the electron DF for various values of $\gamma$ which are greater than unity.
We see that the double maximum in the middle becomes more pronounced as $\gamma$ is increased. This is due to the fact that the Maxwellian $\exp(-\gamma\beta_sH_s)$ multiplying the first part of the DF is now wider than the Maxwellian which multiplies the second part (the $p_{xs}$-dependent population and the first background one are now `colder` than the $p_{ys}$-dependent and second background populations), so the first part dominates and determines the behaviour of the DF. In Figures \ref{fig:gamma1} - \ref{fig:gamma_gt1}, we have chosen the parameters $\bar{a}_0$ and $\bar{b}_0$ such that the DFs are positive for all values of $\gamma$ we consider. As can be seen from the positivity conditions (\ref{a0 bar cond}), the minimum value of $\bar{a}_0$ becomes significantly larger as $\gamma$ is increased (for fixed values of the other parameters). If we were to further increase $\gamma$ then the central `dip` of the DF would become more pronounced, and the DF would become negative, hence we would need to increase $\bar{a}_0$ (and adjust $\bar{b}_0$ if required). %This is why we have only considered values of $\gamma$ that are not much bigger than unity - %, and is significantly large even for $\gamma=5$, due to the factor of $\gamma$ in the exponential in Eq. (\ref{a0 bar cond}). This will lead to a very large maximum density, and hence very small minimum temperature, and so is quite an extreme case, but illustrates nicely the effect of increasing $\gamma$. %For less extreme parameters, the DF tends to be single-peaked in $v_x$.

\subsection{$v_y$-direction}

\begin{figure}[htp]
 \centering{\scalebox{0.4}{\includegraphics{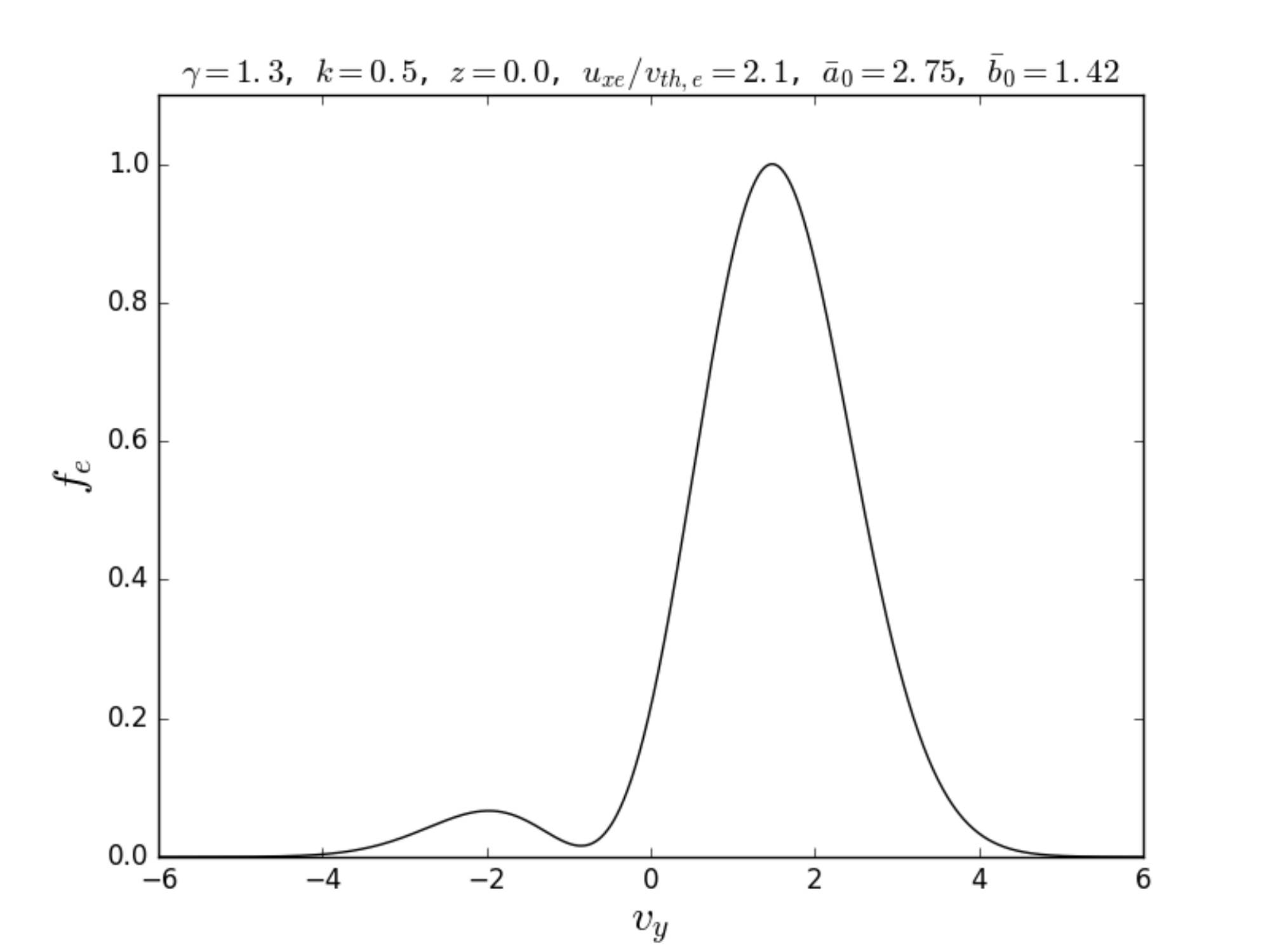}}}
\caption{The electron DF in the $v_y$-direction (with $v_x=v_z=0$) for the parameters used in Figure \ref{fig:gamma_gt1}(c).}\label{fig:vy_gamma1p3_k0pt5}
\end{figure}

In this section we will show some illustrative plots of the electron DF in the $v_y$-direction for various values of $\gamma$. For the parameter set we used in Figures \ref{fig:gamma1} - \ref{fig:gamma_gt1}, the DFs are single peaked in all cases except for $\gamma=1.3$, where there is a double maximum as illustrated in Figure \ref{fig:vy_gamma1p3_k0pt5}. 

From initial investigations, it seems to be difficult to find a set of parameters from which we can illustrate the effect of increasing or decreasing $\gamma$. This may be due to the fact that multiple maxima appear to occur at high values of $u_{xe}/v_{th,e}$, for which we require large values of $\bar{a}_0$ to ensure positivity of the DF - i.e. a large background density. This often results in the DF being single-peaked for smaller values of $\gamma$.

\begin{figure}[htp]
\centering\
\subfigure[]{\scalebox{0.32}{\includegraphics{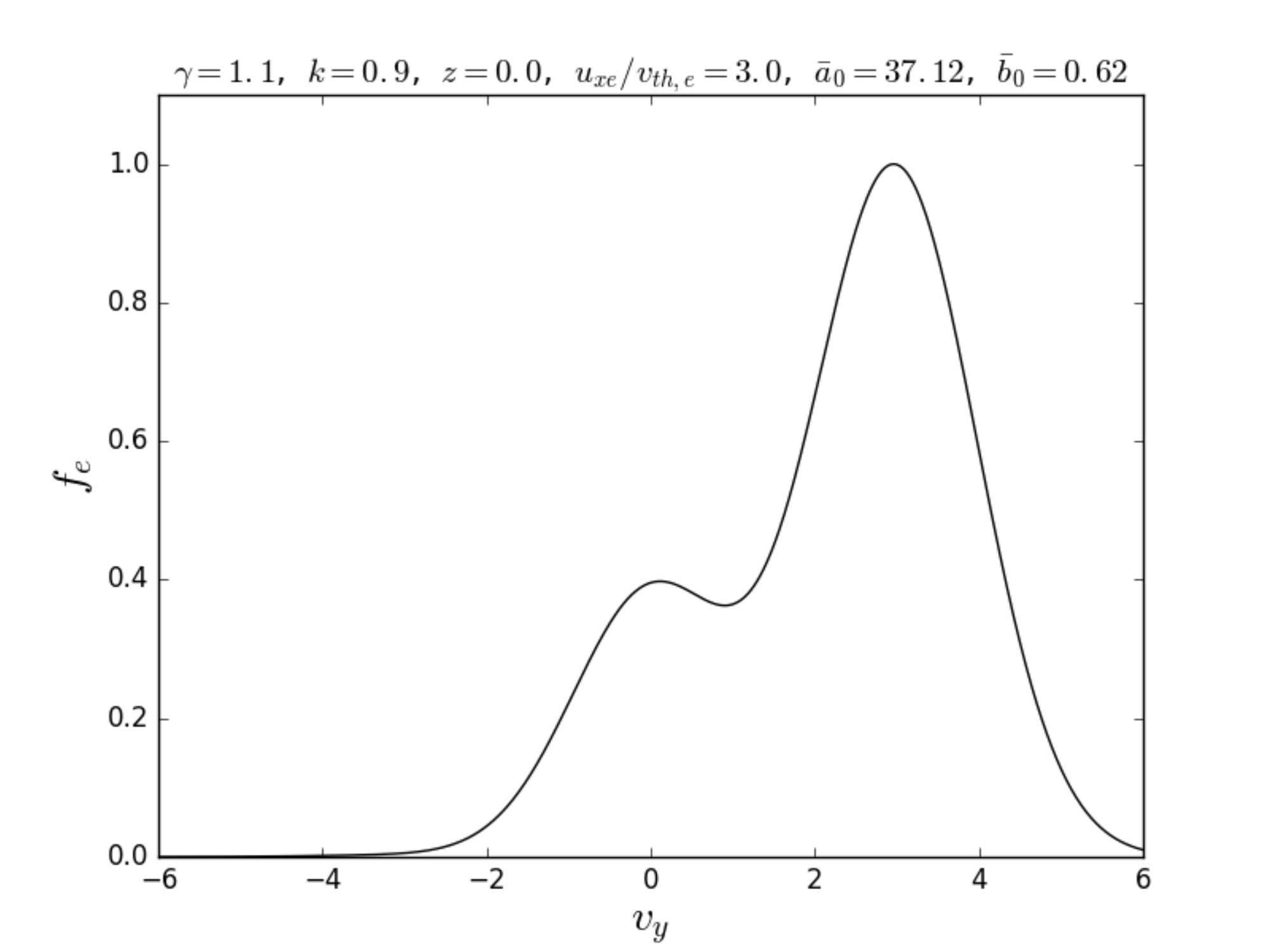}}}
\subfigure[]{\scalebox{0.32}{\includegraphics{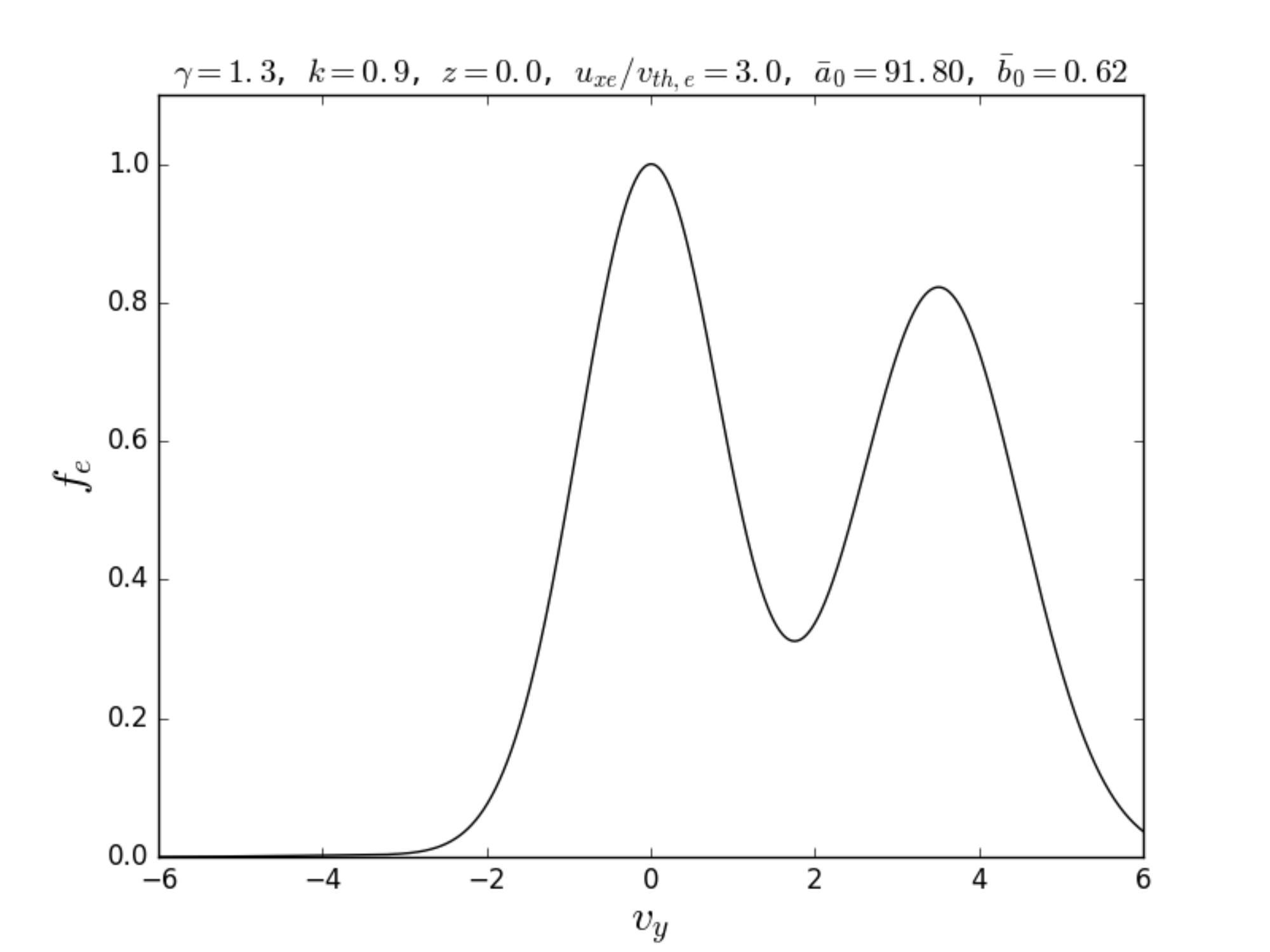}}}
\subfigure[]{\scalebox{0.32}{\includegraphics{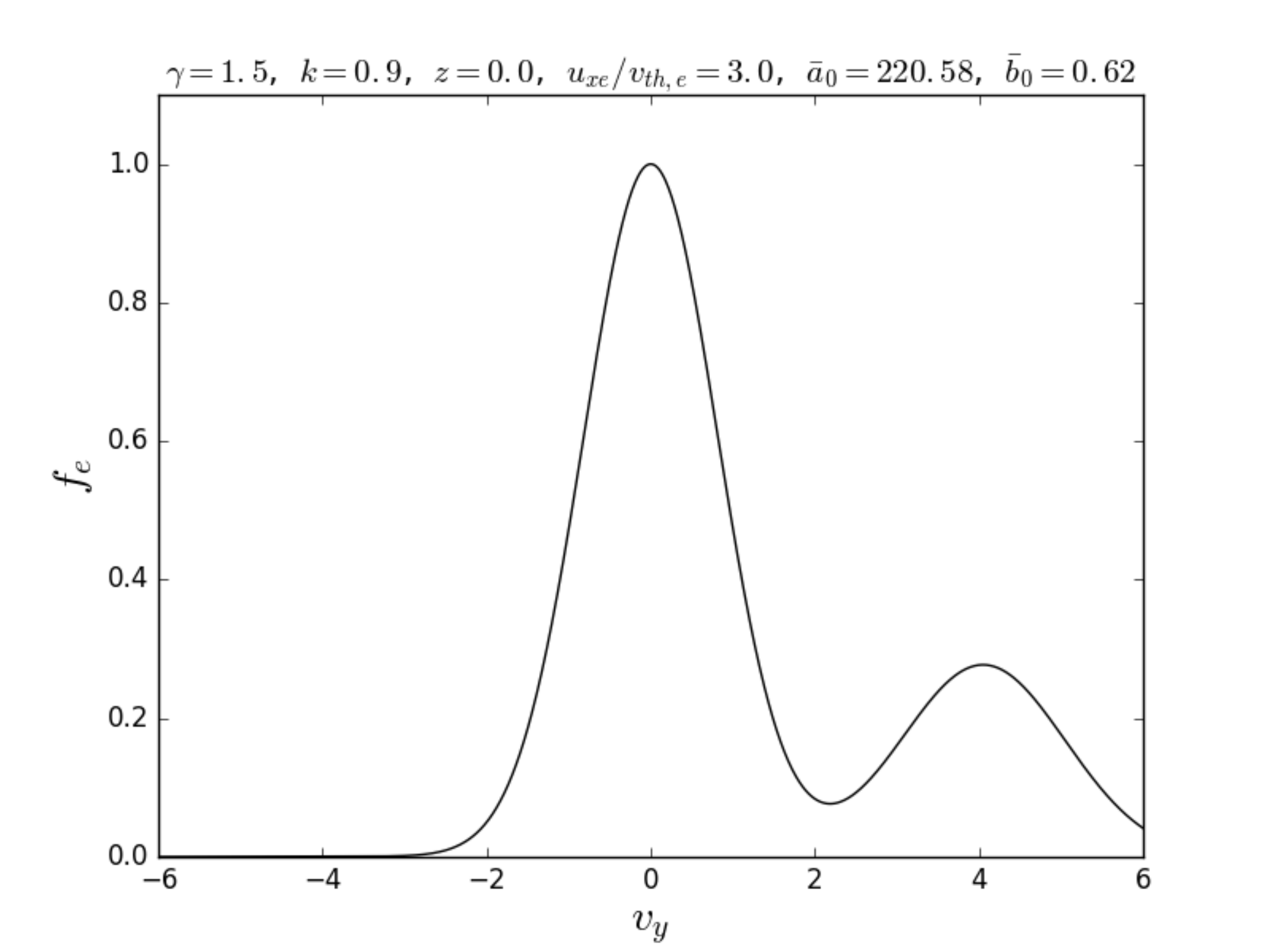}}}
\caption{The electron DF in the $v_y$-direction (with $v_x=v_z=0$) for various parameters sets, to give an illustration of the possible behaviour of the DF in this direction.}\label{fig:vy_examples}
\end{figure}

Possible behaviour of the DF in the $v_y$-direction can be explored heuristically by noting that, for given values $v_x$, $v_z$ and $z$, the DF has the general form
\begin{eqnarray}
f_s(v_y)&=&C_1\exp\left(-\frac{\gamma v_y^2}{2v_{th,s}^2}\right)+C_2\exp\left(-\frac{v_y^2}{2v_{th,s}^2}\right)\nonumber\\
&{}&+C_3\exp\left(-\frac{(v_y+ku_{ys})^2}{2v_{th,s}^2}\right)+C_4\exp\left(-\frac{(v_y-ku_{ys})^2}{2v_{th,s}^2}\right),
\end{eqnarray}
for constants $C_1$-$C_4$, i.e. it consists of two Maxwellian parts with varying widths, and two shifted Maxwellians - one shifted in the positive $v_y$-direction, and the other in the negative $v_y$-direction (by the same amount). Depending on the relative values of $C_1$-$C_4$, therefore, the DF can exhibit different behaviour, some examples of which are given in Figure \ref{fig:vy_examples}. Note that we have taken different values of $\bar{a}_0$ in each plot, to ensure that the DFs are positive in each case.

\section{Summary}
\label{sec:summary}
In this paper, we have presented a class of 1D strictly neutral Vlasov-Maxwell equilibrium DFs for both linear and nonlinear force-free current sheets, with magnetic fields defined in terms of Jacobian elliptic functions, which are an extension of the DFs discussed by \citet{Abraham-Shrauner-2013} to account for non-uniformities in the temperature and density, whilst still maintaining a constant pressure (with respect to the spatial coordinate), as is required for force-balance of the force-free equilibrium. To achieve this, we have used the method of \citet{Kolotkov-2015}, which involves modifying the DF of the original case to include temperature differences between the different particle populations in the model, and then ensuring that strict neutrality is satisfied, and that there is consistency between the microscopic and macroscopic parameters of the equilibrium.

The new DF can be regarded as consisting of four particle populations: one depending on $p_{xs}$, one on $p_{ys}$, and two background populations. The $p_{xs}$-dependent and first background population are taken to have the same energy dependence in the DF, as do both the $p_{ys}$-dependent and second background populations. Note that for the limit of vanishing elliptic modulus, $k$, to give continuous DFs and pressure, density and temperature profiles, we require a particular choice of the constants characterising the background populations, but this form can be changed for other $k$ values if desired (it has the `drawback` of giving a very large maximum density for certain parameter values). 

We have derived sufficient conditions on the parameters such that the positivity of the DFs is ensured, and have given explicit expressions for the density, temperature and pressure across the current sheet. Additionally, we have derived the components of the bulk-flow velocity from the DF, to show that the spatial structure of the current density is determined by the product of the spatial structure of the density and bulk-flow velocity, in contrast to the models of, e.g., \citet{Abraham-Shrauner-2013} and \citet{Neukirch-2009}, where the current density structure is determined solely by the structure of the bulk-flow velocity, and also in contrast to the Harris sheet case \citep{Harris-1962}, where it is determined solely by the density structure.

We have investigated limiting cases of the elliptic modulus, $k$. For $k\to1$ the magnetic field becomes that of the force-free Harris sheet, and in this limit we recover a DF similar to that found by \citet{Kolotkov-2015} for this magnetic field. In the limit $k\to0$, the magnetic field becomes linear force-free, and in Abraham-Shrauner's case the DF takes a form which is similar to one discussed in Refs.~\onlinecite{Channell-1976, Attico-1999,Harrison-2009a}, but which is shifted in $p_{xs}$ and $p_{ys}$. In our extended model, the $k\to0$ limit simply gives an extension of this shifted DF to include non-uniformity in both the temperature and density. %We have also discussed (in an appendix) how we can modify the linear force-free DF discussed by \cite{Channell-1976} (i.e. the 'unshifted' DF). %These linear force-free DFs have not previously been discussed in the literature, and extend the results of \cite{Harrison-2009a}, who show that all DFs of the form $f_s(H_s, p_{xs}^2+p_{ys}^2)$ which give an 'attractive central potential' (i.e. $P_{zz}$) satisfy the VM equations for the linear force-free field referred to above.

We have also illustrated graphically the effect of changing the temperature difference between the particle populations in the DF. In the $v_x$-direction, we found that making the $p_{xs}$ part 'colder' than the $p_{ys}$
 part can result in rather pronounced double maxima of the DF (due to a cosine term in $v_x$), but when the $p_{xs}$ part is 'hotter' these maxima are less significant, or the DF becomes single peaked. In the $v_y$-direction, the DF contains two drifting Maxwellians (with the same energy dependence), and two non-drifting Maxwellians (with different energy dependences), and so there is the possibility of double maxima in the DF depending on the relative values of the coefficients of the separate parts. 
 
Double maxima in the DF may lead to velocity space instabilities (e.g. Ref.~\onlinecite{Gary-book}). Due to the increased complexity of the model, however, we have not attempted a systematic study of the velocity space structure, i.e. we have not derived conditions on the parameters such that the DF can be multi-peaked for some $z$, such has been done by \citet{Neukirch-2009} and \citet{Abraham-Shrauner-2013}. This is left for a future investigation. We note, however, that it will be possible to choose the density of the background populations large enough such that there are only single maxima of the DF over the whole phase space.
%Note, however, that this DF, like Channell's, has the restrictive condition that $T_e/T_i=m_i/m_e$ - \textbf{what does this mean in modified case because temperature now has a different meaning?} The limit $k\to0$ was achieved by redefining constant terms to scale with the modulus $k$ in such a way that the singularities in the DF, density and hence temperature were removed.
\appendix

\section{Parameter relations}
\label{sec:neutrality_app2}
In Section \ref{sec:modified_df}, by imposing the strict neutrality condition $n_e(A_x,A_y)=n_i(A_x,A_y)=n$, we obtain the relations

\begin{align}
 n_{0e}\exp\left(\frac{u_{xe}^2}{2v_{th,e}^2}\right)& = n_{0i}\exp\left(\frac{u_{xi}^2}{2v_{th,i}^2}\right) = n_0\label{modified_neut1}\\
 {a}_{0e}\exp\left(-\frac{u_{xe}^2}{2v_{th,e}^2}\right)& = {a}_{0i}\exp\left(-\frac{u_{xi}^2}{2v_{th,i}^2}\right) = {a}_0\label{a0_cond}\\
 a_{1e}\exp\left(-\frac{(1+\gamma k^2)u_{xe}^2}{2v_{th,e}^2}\right)& = a_{1i}\exp\left(-\frac{(1+\gamma k^2)u_{xi}^2}{2v_{th,i}^2}\right) = a_1,\\
 {b}_{0e}\exp\left(-\frac{u_{xe}^2}{2v_{th,e}^2}\right) & = {b}_{0i}\exp\left(-\frac{u_{xi}^2}{2v_{th,i}^2}\right) = {b}_0.\label{b cond}\\
 b_{1e}\exp\left(\frac{k^2u_{ye}^2-u_{xe}^2}{2v_{th,e}^2}\right)& = b_{1i}\exp\left(\frac{k^2u_{yi}^2-u_{xi}^2}{2v_{th,i}^2}\right) = b_1\\
b_{2e}\exp\left(\frac{k^2u_{ye}^2-u_{xe}^2}{2v_{th,e}^2}\right)& = b_{2i}\exp\left(\frac{k^2u_{yi}^2-u_{xi}^2}{2v_{th,i}^2}\right) = b_2\\
\beta_e\vert u_{xe}\vert&=\beta_i\vert u_{xi}\vert\label{uxiuxe}\\
-\beta_eu_{ye}&= \beta_iu_{yi}.\label{modified_neut2}
\end{align}
%Instead of imposing the conditions (\ref{a0_cond}) and (\ref{b cond}), it might seem natural to just let $(a_{0s}+b_{0s})\exp(-{u_{xs}^2}/({2v_{th,s}^2}))$ be equal for ions and electrons, but we have not done thisbecause this would cause problems when trying to write the pressure in terms of the species-independent parameters (due to factors of $\gamma^{-1}$ appearing in parts of the expression). 
Using the choices (\ref{a0s choice}) and (\ref{b0s choice}) for $a_{0s}$ and $b_{0s}$, the conditions (\ref{a0_cond}) and (\ref{b cond}) can equivalently be written as 
\begin{align}
 \bar{a}_{0e}\exp\left(-\frac{u_{xe}^2}{2v_{th,e}^2}\right) &= \bar{a}_{0i}\exp\left(-\frac{u_{xi}^2}{2v_{th,i}^2}\right) = \bar{a}_0\\
 \bar{b}_{0e}\exp\left(-\frac{u_{xe}^2}{2v_{th,e}^2}\right) &= \bar{b}_{0i}\exp\left(-\frac{u_{xi}^2}{2v_{th,i}^2}\right) = \bar{b}_0,
\end{align}
where $a_0=\bar{a_0}+\gamma/(2k^2)$, $b_0=\bar{b}_0-1/(2k^2)$.

By calculating two expressions for the pressure $P_{zz}$, in terms of the macroscopic and microscopic parameters of the equilibrium respectively, and comparing these expressions, we obtain the relations

\begin{eqnarray}
 n_0\frac{\beta_e+\beta_i}{\beta_e\beta_i}&=&\frac{B_0^2}{2\mu_0}\label{micromacro1}\\
  \frac{{a}_0}{\gamma}+{b}_0&=&\frac{P_{t1}+P_{t2}}{B_0^2/2\mu_0}-\frac{3}{2}\label{p mm}\\
 \frac{a_1}{\gamma}&=&-\frac{1}{2k^2}\label{a1}\\
 b_1&=&\frac{1}{4}\left(\frac{1}{k}+1\right)^2\label{a2}\\
 b_2&=&\frac{1}{4}\left(\frac{1}{k}-1\right)^2\label{a3}\\
 \frac{2}{B_0L}&=&\gamma\beta_s\vert u_{xs}\vert q_s=\beta_su_{ys}q_s\Rightarrow u_{ys}=\gamma \vert u_{xs}\vert.\label{ux uy gamma}
\end{eqnarray}
Similarly to previous work (e.g. Ref.~\onlinecite{Neukirch-2009}), we can derive an expression for the current sheet half-width $L$, in terms of the microscopic parameters, as
\begin{equation}
L=\left({\frac{2(\beta_e+\beta_i)}{\mu_0e^2\beta_e\beta_in_0(u_{yi}-u_{ye})^2}}\right)^{1/2}.\label{asL}
\end{equation}

\acknowledgments{We acknowledge the support of the Science and Technology Facilities Council via the consolidated grants ST/K000950/1 and ST/N000609/1 and the doctoral training grant ST/K502327/1 (O. A.), and the Natural Environment Research Council via grant no. NE/P017274/1 (Rad-Sat) (O. A.). F. W. and T. N. would also like to thank the University of St Andrews for general financial support.}

%\clearpage
%\FloatBarrier
%\bibliographystyle{unsrtnat}
%\bibliography{fiona}

\end{document}